\documentclass[12pt,letterpaper]{article}

\usepackage{graphicx}
\usepackage{color}
\usepackage[body={16.5cm, 22cm},right=2.6cm]{geometry}
\usepackage{amssymb}
\usepackage{amsmath} 
\usepackage{epsfig}
\usepackage{psfrag}

\newcommand\ee{\end{equation}}
\newcommand\be{\begin{equation}}
\newcommand\eea{\end{eqnarray}}
\newcommand\bea{\begin{eqnarray}}
\newcommand{\sfrac}[2]{{\textstyle\frac{#1}{#2}}}
\newcommand\di{\partial}

\def\e{{\rm e}}

\def\d{\partial}
\def\l{\left(}
\def\r{\right)}
\def\beq{\begin{equation}}
\def\eeq{\end{equation}}
\def\de{\partial}
\def\la{\langle}
\def\ra{\rangle}

\def\nn{\nonumber}
\def\barr{\begin{array}}
\def\earr{\end{array}}
\def\Re{{\rm Re}}

\begin{document}

\begin{titlepage}
\begin{flushright}
\end{flushright}
\vskip 1.5cm
\begin{center}


{\huge \sc Quantum Horizons of the \\[0.4cm] 
Standard Model Landscape }\\[1cm]
{\large  \bf Nima Arkani-Hamed$^{a}$,  
Sergei Dubovsky$^{a,b}$, \\[.3cm]
Alberto Nicolis$^{a}$ and  
Giovanni Villadoro$^{a}$}
\\[.66cm]
{\normalsize { \sl $^{a}$ Jefferson Physical Laboratory, \\ Harvard University, Cambridge, MA 02138, USA}}

\vspace{.2cm}
{\normalsize { \sl $^{b}$ Institute for Nuclear Research of the Russian Academy of Sciences, \\
        60th October Anniversary Prospect, 7a, 117312 Moscow, Russia}}

\vskip1.4cm {\bf Abstract\\[10pt]} 
\end{center}
 {\small
The long-distance effective field theory of our Universe---the Standard 
Model coupled to gravity---has a unique 4D vacuum, but we show that it also 
has a landscape of lower-dimensional vacua, with the potential for moduli 
arising from vacuum and Casimir energies. For minimal Majorana neutrino 
masses, we find a near-continuous infinity of AdS$_3 \times$S$_1$ vacua, 
with circumference $\sim 20$ microns and AdS$_3$ length $4 \times 10^{25}$~m.
By AdS/CFT, there is a CFT$_2$ of central charge $c \sim 10^{90}$ which contains 
the Standard Model (and beyond) coupled to quantum gravity in this vacuum.
Physics in these vacua is the same as in ours for energies between 
$10^{-1}$~eV and $10^{48}$~GeV, so this CFT$_2$ also describes all the physics of our vacuum in this energy range. 
We show that it is possible to realize 
quantum-stabilized AdS vacua as near-horizon regions of new kinds of 
quantum extremal black objects in the higher-dimensional space---near critical black 
strings in 4D, near-critical black holes in 3D. The violation of the null-energy condition by the 
Casimir energy is crucial for these horizons to exist, as has already been 
realized for analogous non-extremal 3D black holes by Emparan, Fabbri 
and Kaloper. The new extremal 3D black holes are particularly 
interesting---they are (meta)stable with an entropy independent of 
$\hslash$ and $G_N$, so a microscopic counting of the entropy may be possible in the $G_N \to 0$ limit. 
Our results suggest that it should be possible to realize the larger landscape of 
AdS vacua in string theory as near-horizon geometries of new extremal black brane solutions.  
}

\vspace{4cm}

%
\end{titlepage}

\tableofcontents


\section{Preamble}

M-theory is a unique theory with a unique 11-dimensional vacuum. However it 
also has an enormous landscape of lower-dimensional vacua, which raises 
the thorny questions of vacuum selection. The long distance effective 
theory of our world---the Standard Model coupled to gravity---is an 
effective field theory in 4 dimensions, with some fixed microphysics, and 
also has a unique 4D vacuum. In this paper, we begin by showing that there 
is also a Standard Model landscape, by exhibiting a near-continuous 
infinity of lower-dimensional vacua of the theory. The simplest example is 
compactification on a circle, where the potential for the radius modulus 
receives competing contributions from the tiny cosmological constant, as 
well as the Casimir energies of the graviton, photon and, crucially, the 
massive neutrinos. With Majorana neutrinos, whose masses are constrained by 
explaining the atmospheric and solar neutrino anomalies, we find an AdS$_3\times$S$_1$ 
vacuum of the theory with the circumference of the S$_1$ at 
about $\sim 20$ microns. With Dirac neutrinos, both AdS$_3$ as well as 
dS$_3$ vacua are possible. Lower-dimensional vacua can exist as 
well. These solutions exist completely independently of any UV completion 
of the theory at the electroweak scale and beyond.  Of course if string 
theory is correct and our 4D vacuum is part of the theory, then the 
vacua we are describing are part of the string landscape as well. While 
we focus on the Standard Model landscape here, such vacua would seem 
to be generic in non-supersymmetric theories where the cosmological 
constant is fine-tuned to be small.

The AdS$_3$ vacua are particularly interesting: it is often thought that 
AdS/CFT can not be used to describe quantum gravity in our world because 
we have a positive cosmological constant. But this is not the 
case in these AdS$_3 \times$S$_1$ vacua! By AdS/CFT, there must be some 
two-dimensional conformal field theory description of this background. 
Since the size of the S$_1$ is so large, all of 
conventional high-energy physics---the spectrum of leptons and hadrons, 
electroweak symmetry breaking, whatever completes the Standard Model up to 
the Planck scale, even very high energy scattering probing quantum gravity 
at energies well above the Planck scale but beneath energies that would 
make a $\sim 20$ micron sized black-hole---is the same in this vacuum as 
in ours. Of course we can't yet identify this CFT, but it's existence as 
the dual description of quantum gravity in a {\it very} close cousin of 
our world is quite interesting.

After discussing the vacua, we turn to the interesting question of what 
physical processes can connect or interpolate between them. We will see 
that there are novel extremal black holes and black strings which
asymptote to the 4D vacua and realize the lower-dimensional AdS vacua as 
their near-horizon geometries, in a way analogous to ordinary extremal 
charged black holes and branes that interpolate from Minkowski space to 
AdS$_m \times$S$_n$ vacua. What is interesting is that these are 
intrinsically {\it quantum} black objects---such horizons can not exist 
classically due to familiar no-hair arguments which follow from 
an energy momentum tensor satisfying the null energy condition. However 
the energy conditions are violated by the Casimir energies, which play the 
crucial role in modulus stabilization to begin with. Schwarzschild-type 
non-extremal quantum black holes supported by Casimir energy have been 
studied recently by Emparan, Fabbri and Kaloper~\cite{Emparan:2002px}; our further 
contribution here is (A) to realize that these objects exist as solutions 
in the Standard Model and (B) to place them in a broader context, revealing 
also the extremal black holes and their role as interpolators in the 
Standard Model landscape. These novel sorts of black hole are very 
interesting and we will discuss a number of their properties. We will also 
discuss some interpolations to the lower-dimensional dS vacua as well.

In the simplest case of the Standard Model AdS$_3\times$S$_1$ vacuum
 the interpolating solutions are cosmic strings.  Smallness of the Casimir
 potential implies that the opening angle is very small, so that such cosmic string cannot be present in
 the visible part of the Universe. However, given that we live in de Sitter space, there is a (tiny) non-zero probability for a dS thermal fluctuation
resulting in the creation of this object within our causal patch.
Note that this transition does not change the microscopic structure of the
vacuum at distances smaller than 20 microns, so that small enough observers---for instance, many of the {\it  Amoebozoa}---are able to  survive it and enter the lower-dimensional vacua.

It is interesting that the presence of some ``negative" gravitational 
energy, violating the null-energy condition, is a crucial part of {\it all} 
realistic modulus stabilization mechanisms; in string theory a common source of 
the negative energies come from the negative-tension orientifold planes, 
while in our Standard Model vacua it arises from Casimir energy. It is 
natural to conjecture that {\it all} the AdS vacua in the larger string 
landscape can be thought of as the near-horizon limits of ``exotic" 
extremal black holes in 10 dimensions, with the no-hair theorems being 
evaded by the negative energies needed for modulus stabilization. If true, it would be interesting to probe these vacua from the 
``outside".

\section{The Standard Model Landscape}
\label{sec:SMlandscape}

We will now show that the action  of the minimal
Standard Model (SM) plus General Relativity (GR) has more than one distinct vacuum,
actually a true landscape of vacua.
Let us start considering the SM+GR action compactified on a circle of radius $R$. 
At distances larger than $R$, there is an effective 3D theory with a metric
parameterized by
\beq \label{eq:4to3metric}
ds_{(4)}^2=\frac{r^{2}}{R^2} ds_{(3)}^2+R^{2} \l d \phi- \frac{\sqrt2}{M_4\, r} V_{\mu} dx^\mu\r^2\,,
\eeq
where $M_4$ is the 4D reduced Planck mass ($1/\sqrt{8\pi G_N}$), 
$R$ is the radion field, $V_{\mu}$ is the graviphoton, $\phi\in[0,2\pi)$,
and $r$ is an arbitrary scale that we will later fix to the expectation value of $R$.
With such parameterization the effective action is already in the Einstein frame,
in particular, the reduction of the action for the pure gravitational sector reads
\bea
S_{\rm grav}&=&\int d^3x\,d\phi \sqrt{-g_{(4)}} \l \frac12 M_4^2\, {\cal R}_{(4)}- \Lambda_4 \r  \nn \\ 
&\to&\int d^3x \sqrt{-g_{(3)}} \, (2\pi r) \left[ \frac12 M_4^2\, {\cal R}_{(3)}-\frac14 \frac{R^4}{r^4} \, V_{\mu\nu} V^{\mu\nu}
-M_4^2\l\frac{\de R}{R}\r^2-\frac{r^2\Lambda_4}{R^2}\right] , \nn
\eea
where  $\Lambda_4$ is the
4D cosmological constant and $V_{\mu\nu}$ is the field strength of the graviphoton.
Because of the 4D cosmological constant, the classical potential for the radion
is {\it runaway}, which makes the circle decompactify. 
Indeed this rolling solution is just the expanding 4D de Sitter solution.

However, the smallness of the cosmological constant and the absence of other classical
contributions to the effective potential for the radion make quantum corrections 
important for the study of the stabilization of the compact dimension.
The 1-loop corrections to the radion potential is the Casimir energy coming
from loops wrapping the circle, which are UV insensitive and calculable.
The Casimir potential for a particle of mass $m$ is $\propto e^{-2\pi m R}$ for $R\gg 1/m$,
so at any scale $R$, only particles with mass lighter than $1/R$ are relevant.
The contribution to the effective potential of a massless state (with periodic boundary conditions) 
is
\beq \label{eq:masslesscas}
\mp\frac{n_0}{720 \pi}\frac{r^3}{R^6}\,,
\eeq
where the sign $\mp$ is for bosons/fermions and $n_0$ is the number of degrees of freedom
(see Appendix~\ref{app:Casimir} for details). The only massless particles (we know of!) in the SM
are the graviton and the photon. For very large radii the cosmological constant 
contribution wins and the radion potential is runaway
while for small radii the Casimir force wins and the compact dimension start shrinking.
We thus get a maximum for $R=R_{max}$, with
\beq \label{eq:maximum}
R_{max}=\l\frac{1}{120 \pi^2 \Lambda_4} \r^{\frac14}\,,
\eeq
where we put $n_{0}=4$ in eq.~(\ref{eq:masslesscas}) (2 from the graviton + 2 from the photon) and which,
for the current value of the cosmological constant $\Lambda_4\simeq 3.25\cdot10^{-47}$~GeV$^4$~\cite{pdg},
means $R_{max}\simeq 14$~microns.

\begin{figure}[t]
\psfrag{V}{\hspace{-16pt} $V(R)$} \psfrag{cc}{$\Lambda_4$} \psfrag{bos}{$g,\gamma$} \psfrag{nu}{$\nu_i$} \psfrag{R}{$R$} 
\psfrag{ads}{\footnotesize $m_\nu^4 >\Lambda_4$} 
\psfrag{dS}{\footnotesize $m_\nu^4 \sim\Lambda_4$} 
\psfrag{none}{\footnotesize $m_\nu^4 <\Lambda_4$} 
\begin{center} \epsfig{file=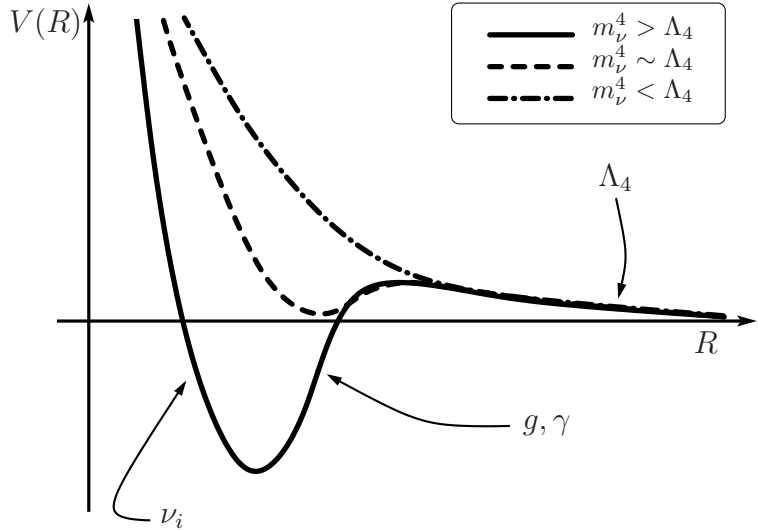,width=0.58\textwidth} \end{center} 
\vspace{-0.6cm}
\caption{{\it \small Radion potential around the neutrino-cosmological constant scale.
The regions where the cosmological constant, the massless bosons and the neutrino
contributions dominate are indicated with arrows. Depending on the neutrino spectrum
the three plots show the three possible scenarios: no vacua, dS and AdS vacuum.}}
\label{fig:neutrino-vacuum}
\end{figure}

If we start with a size $R$ smaller than this critical value, the circle wants to shrink,
however, when the inverse size becomes comparable to the lightest massive particle, its contribution
to the effective potential is not suppressed anymore and can change the behavior of the potential.
This is indeed what happens when $1/R$ approaches the neutrino mass scale.
The contribution of fermions to the Casimir energy is indeed opposite to that of bosons
and since the neutrino d.o.f. are at least 6 (for Majorana neutrino, 12 for Dirac) at shorter
scales their contribution eventually wins against that of bosons. 
Thus a local minimum in general appears.
However, since neutrino masses are of the same order as the scale (\ref{eq:maximum}), the actual existence
of the minimum can depend on the details of the neutrino mass spectrum (see Fig.~\ref{fig:neutrino-vacuum}).

On S$_1$ there is a discrete choice for the spin connection, which results in the choice of
periodic or antiperiodic boundary conditions for fermions. In the first case the contribution
has opposite sign with respect to that of bosons, while in the second case is the same. 
In order to have a minimum we thus need to impose periodic boundary conditions for the neutrinos
and have no more than 3 light fermionic d.o.f., where light here means lighter than
the scale of eq.~(\ref{eq:maximum}). If these conditions are not met, the positive contributions
from the neutrinos start overwhelming the bosonic ones before the latter are able to develop a maximum,
and no minimum is developed as well. 

We do not know yet the actual neutrino spectrum, nor whether neutrinos are Majorana or Dirac particles.
We only know the mass splittings for solar and atmospheric neutrino oscillations~\cite{pdg},
\bea
\Delta m_{{\rm atm}}^2 &\simeq& (1.9\div 3.0) \cdot 10^{-3}\, {\rm eV}^2\,,\nn \\
\Delta m_{\odot}^2 &\simeq& (8.0\pm0.5) \cdot 10^{-5}\, {\rm eV}^2\,. \label{eq:deltam}
\eea
If we call $\nu_{i}$, with $i=1\dots3$, the $i$-th mass eigenstate, such that 
$m_{\nu_1}<m_{\nu_2}<m_{\nu_3}$, we have two possibilities: (a) the normal hierarchy spectrum with
$\Delta m_{12}^2=\Delta m_{\odot}^2$, $\Delta m_{23}^2=\Delta m_{{\rm atm}}^2$, (b) the inverted hierarchy
spectrum with $\Delta m_{12}^2=\Delta m_{{\rm atm}}^2$, $\Delta m_{23}^2=\Delta m_{\odot}^2$.
From eq.~(\ref{eq:deltam}) it follows that even assuming $m_{\nu_1}=0$, independently of the hierarchy structure of the neutrino 
mass spectrum, $m_{\nu_2}\gtrsim 9\cdot10^{-3}$~eV.

\begin{figure}[t]
\psfrag{xx}{}
\psfrag{X}{\hspace{-20pt} \footnotesize $R$ (GeV$^{-1}$)}
\psfrag{Y}{\hspace{-20pt} \footnotesize $V(R)$ (GeV$^3$)}
\begin{center}
\begin{tabular}{cc}
\epsfig{file=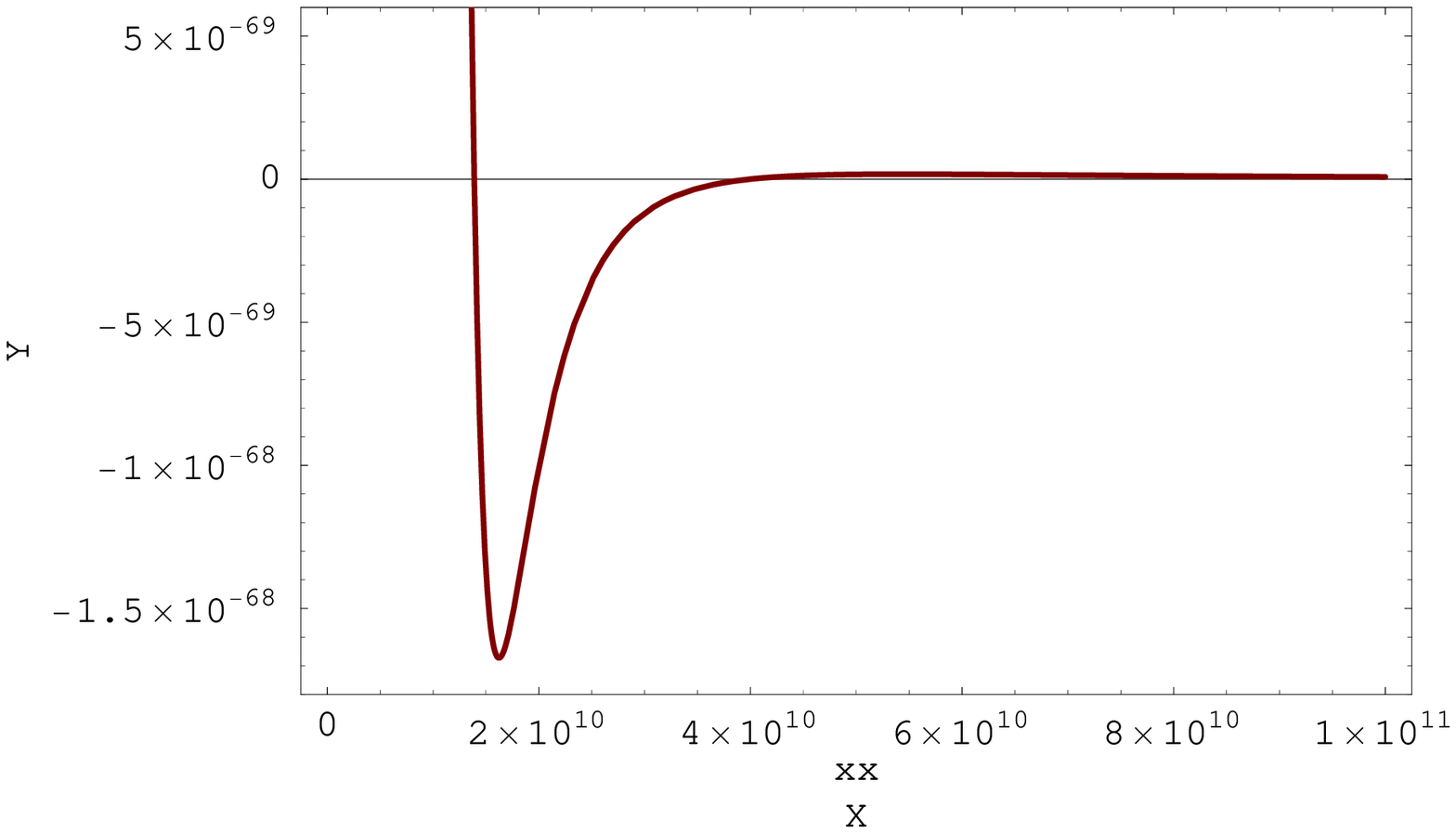,height=5cm} &
\raisebox{0.2cm}[0pt][0pt]{\epsfig{file=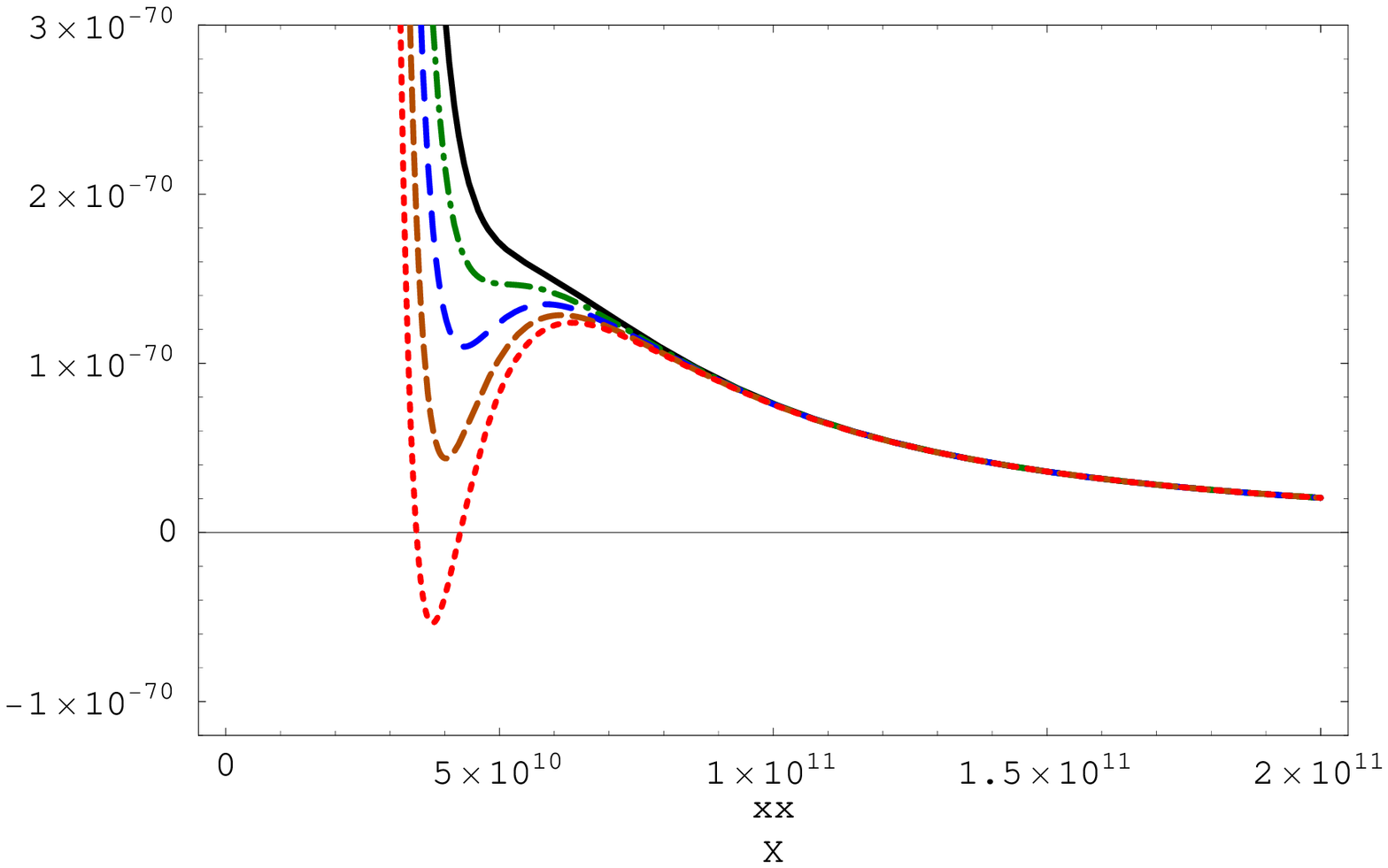,height=4.75cm}} \\
\footnotesize a) & \footnotesize b)
\end{tabular}
\end{center}
\vspace{-0.6cm}
\caption{{\small \it Effective potential as a function of the radion field $R$, in the case of {\rm a)}~Majorana 
and {\rm b)}~Dirac neutrinos for different choices of the lightest neutrino mass: $m_{\nu_1}$=$0$ for Majorana,  
$m_{\nu_1}$=$6.5$ (continuous black), $7.0$ (dash-dotted green),
$7.5$ (long dashed blue), $8.0$ (short dashed brown) and $8.5$ (dotted red) {\rm meV}, for Dirac neutrinos.
The other neutrino masses (with normal hierarchy) have been fixed with the current values for the mass splittings of eq.~(\ref{eq:deltam}).
In the plot the scale $r$ has been chosen so that $2\pi r$=$1$~{\rm GeV}$^{-1}$.}}
\label{fig:majorana-dirac-vacuum}
\end{figure}

If neutrinos are Majorana particle, this means that no more than 2 d.o.f. can be lighter than $1/(2\pi R_{max})$. 
In this case we have necessarily a new Standard Model vacuum! 
The effective potential at this minimum is always negative (Fig.~\ref{fig:majorana-dirac-vacuum}a),
therefore this vacuum solution is AdS$_3\times$S$_1$. The radion at the minimum ($R_0$) is of order $1/m_{\nu}$,
while both the AdS$_3$ length ($\ell_{3}$) and the radion mass ($m_R$) are of order of the 4D Hubble scale ($\ell_4$).
Just to give some numbers, if, for example, we take $m_{\nu_1}=0$, 
$m_{\nu_2}^2=\Delta m_{\odot}^2$ and $m_{\nu_3}^2=\Delta m_{{\rm atm}}^2$, we have
\bea
R_{0}&\simeq& 3.2\, {\rm \mu m}\,,  \nn \\
\ell_{3} &\simeq& 4\, \ell_{4} \simeq 3.7\cdot 10^{25}\, {\rm m} \,,\nn \\
m_R&\simeq& 6.5/\ell_{3} \simeq 3.5\cdot 10^{-41}\, {\rm GeV}\,. \nn
\eea

If on the other hand, neutrinos are Dirac, then from eq.~(\ref{eq:deltam}) we get an AdS$_3$ minimum only if
the lighter neutrino mass $m_{\nu_1}$ is larger than $\approx 8.3\cdot 10^{-3}$~eV (normal hierarchy)
or $\approx 3.1\cdot 10^{-3}$~eV (inverted hierarchy), a metastable 
dS$_3$ minimum if $m_{\nu_1}\approx (7.1\div8.3)\cdot 10^{-3}$~eV (normal hierarchy) or 
$m_{\nu_1}\approx (2.5\div3.1)\cdot 10^{-3}$~eV (inverted hierarchy), and no stationary point if 
$m_{\nu_1}\lesssim 7.1\cdot 10^{-3}$~eV (normal hierarchy) or 
$m_{\nu_1}\lesssim 2.5\cdot 10^{-3}$~eV (inverted hierarchy), see Fig.~\ref{fig:majorana-dirac-vacuum}b.

Depending on the neutrino vacua we can thus have a 3D vacuum with positive,
zero or negative cosmological constant. In either case the natural value for the effective vacuum 
energy will be
\beq \label{eq:L3nat}
\Lambda_3\sim m_{\nu}^3\approx \Lambda_4 R_0\,.
\eeq
In the case of positive $\Lambda_3$ we have a 3D dS vacuum. It is interesting 
to compare the entropy $S_3$ associated to the dS$_3$ horizon 
with the 4D one ($S_4$). We thus have
\beq
S_3=\frac{M_3}{H_3}\sim M_4^3 R_0^3\approx\frac{m_\nu}{M_4} S_4\,, \nn
\eeq
which is much smaller than the 4D dS entropy.
In principle, one could also have $S_3>S_4$, since in the limit $\Lambda_3\to0$
$S_3\to\infty$, however, one would need $\Lambda_3$ to be suppressed with respect to
its natural value in eq.~(\ref{eq:L3nat}) by a factor of $m_\nu^2/M_4^2$, which
turns into a $10^{-60}$ tuning on the neutrino masses.
\begin{figure}
\psfrag{xx}{}
\psfrag{X}{\hspace{-20pt} \footnotesize $R$ (GeV$^{-1}$)}
\psfrag{Y}{\hspace{-20pt} \footnotesize $V(R)$ (GeV$^3$)}
\begin{center}
\epsfig{file=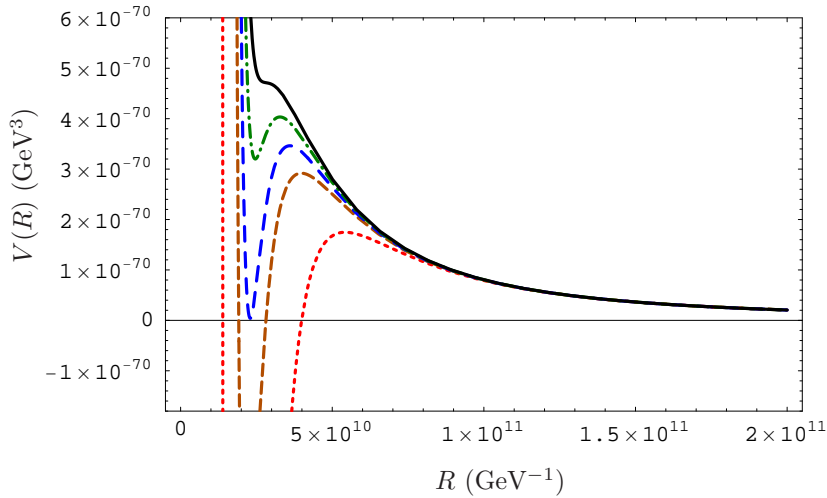,width=0.66\textwidth}
\end{center}
\vspace{-0.6cm} 
\caption{{\it \small  Sensitivity of the Majorana neutrino vacuum to $\Delta m_\odot^2$.
The plot shows the radion potential for different values of $\Delta m_{12}^2=\Delta m_\odot^2$:
$\Delta m_\odot^2$=$8.0$ (dotted red, current value), $2.0$ (short dashed brown),
$1.5$ (long dashed blue), $1.2$ (dot-dashed green) and $1.0$ (continuous black) $\cdot\, 10^{-5}$~{\rm eV}.
The lightest neutrino mass has been fixed to zero. In the plot the scale $r$ 
has been chosen so that $2\pi\,r$=$1$~{\rm GeV}$^{-1}$.}}\label{fig:deltams}
\end{figure}

Let us stress again that the above analysis depends entirely on IR physics
and is independent of UV details, indeed
the first non vanishing corrections would come from the electron (the next lightest state)
whose contribution is suppressed by $e^{-2\pi m_e R_0}\sim e^{-m_e/m_\nu}$! 
The calculation is also stable with respect to higher order quantum corrections, which are
small as long as the 4D couplings are perturbative.

The presence and the properties of the neutrino vacuum are very sensitive to both the value 
of the cosmological constant and the neutrino spectra (see Fig.~\ref{fig:majorana-dirac-vacuum} and \ref{fig:deltams}). 
If the cosmological constant had been natural, of order of the Planck or some other high scale, 
quantum effects would have been negligible at low energies and would have not
been able to produce any vacuum.
Indeed, just increasing the value of the cosmological constant by an order of magnitude,
would be enough to eliminate the presence of the 3D vacuum.  
Analogously, as shown in Fig.~\ref{fig:deltams}, it is enough to decrease the mass of
the second lightest neutrino just by a factor of 3, which means a factor 9 in $\Delta m_\odot^2$, 
to destroy the Majorana-neutrino vacuum (for normal hierarchy).
%

\subsection{A near moduli space}
Besides the radion there is another modulus in the compactified 3D action:
the longitudinal polarization of the photon $A_\phi$. Classically, because of gauge invariance,
this field is massless. However, at the quantum level, the mass of this field in general gets corrections 
that depend on the gauge invariant (Wilson loop) combination:
\beq
W=\exp\l i\oint_{S_1} A\r \,. \nn
\eeq
These corrections are generated at one loop by charged fields wrapping S$_1$.
They can be easily calculated by noticing that, via a gauge transformation,
the Wilson loop can be reabsorbed into a non-trivial boundary condition for the charged field.
Since a change of the boundary condition will change the contribution of a field
to the energy density, this produce a non trivial potential also for $A_\phi$.
The explicit formula for the potential for $R$  and $A_\phi$ can be found in the Appendix~\ref{app:Casimir} (eq.~(\ref{eq:Vpot})).

In general charged fermions want to stabilize the Wilson loop around $A_\phi=1/2$ while charged 
bosons around $A_\phi=0$.
In our case the first contribution comes from the electron, it produces a cosine-like 
potential that stabilizes the Wilson loop around $A_\phi=1/2$.
There are thus two stationary points, a minimum around $A_\phi=1/2$ and a maximum around $A_\phi=0$.
According to \cite{bf} both are stable in AdS.

However, because the electron is very heavy compared to our scale,
its contribution is exponentially suppressed by a factor of order $e^{-m_e/m_\nu}\sim e^{-10^8}$!
Because of the smallness of this contribution one may worry whether other would-be subleading 
corrections may be important. For instance, at higher loop order there are 
contributions to the effective action with the photon going around the loop. Such 
corrections are power like and, although subleading in $\alpha_{em}$, might
nevertheless be important. It is easy to show, however, that such corrections do not
generate a potential for $A_\phi$ at any order in perturbation theory. Indeed, as long
as $R\gg m_e^{-1}$, we can integrate out the electron and use the Euler-Heisenberg effective action.
In this effective theory there are no minimally coupled particles,
all fields are gauge invariant so that the photon possesses an exact shift symmetry that protect it
from mass terms at all order in the energy expansion. The perturbative
expansion is accurate up to non-perturbative corrections in $E/m_e$ of order $e^{-m_e/E}\sim e^{-2\pi\, m_e R_0}$,
which is of the same order of the contribution from the electron. 

Therefore the potential for $A_\phi$ is effectively flat, in the sense that starting from any 
value of $W$ it would take an exponentially long time (say in the AdS$_3$ length units) 
to move to the minimum.
In this sense the neutrino-vacuum is not unique, effectively there is a continuum of distinct
vacua labeled by different values of $W$.
Strikingly enough we see that the Standard Model, although non-supersymmetric, 
possesses a near moduli space.
The phenomenon that a unique action may give rise to an infinite number of vacua is not
a special feature of Superstring/SUSY theories, it is also a feature of the minimal Standard Model!

\subsection{More Vacua}
Our analysis in the previous sections was restricted to the simplest 
Standard Model on a micron sized circle. However, it is natural
to expect that there are more vacua. For instance, for smaller
size of the circle, more SM states start contributing to the Casimir energy,
when bosons and fermions contributions compensate each other, the radion potential can develop
a stationary point. The analysis is reported in Appendix~\ref{app:more3Dvacua}; apart
for a saddle point at the electron scale no new stationary point is present until
$R \sim \Lambda_{QCD}^{-1}$. 
The study of the radion potential around the QCD scale would require a non-perturbative
analysis. Above this scale, the theory becomes perturbative again. We give
the general formula for the effective potential in Appendix~\ref{app:more3Dvacua}
but we do not attempt to address the stabilization problem since now the structure of 
the potential is complicated by the presence of more Wilson loop moduli 
from gluons and at still shorter distances from electroweak bosons.
Extensions of the Standard Model can also affect the Casimir potential, creating new vacua
or removing the existing ones. 
For example the presence of light bosonic fields, 
like the QCD axion, or extra-dimensional light moduli may favor the presence of
the micron vacuum in the case of Dirac neutrinos, while very light fermions, like
goldstinos, gravitinos or sterile neutrinos, would tend to destroy such vacuum. 
Another example is supersymmetry at the TeV scale, which would even the number of
bosonic and fermionic d.o.f. and give room to the presence of new vacua at that scale.
We comment on some of these possibilities in the Appendix~\ref{app:ext3Dvacua}.

Another possibility is to compactify more than one dimension.
We summarize here the main features of such lower-dimensional vacua, 
and refer the reader to the Appendix (\ref{2Dvacua} and \ref{1Dvacua}) for a detailed analysis.
If we compactify two of the spatial dimensions, at low energies the system is well 
described by a 2D effective theory containing gravity and a set of scalar fields that parameterize 
the overall size and shape of the manifold we are compactifying on. For instance if we compactify on a two-torus, 
beside gravity the 2D theory contains the area field $A$ and the complex modulus $\tau = \tau_1 + i \tau_2$.
The analysis of this system is subtler than in the usual case of toroidal compactifications in higher-dimensional models, 
for in our low-dimensional setups several degrees of freedom are not dynamical. 
Gravity itself is not dynamical in 2D, neither is the area $A$. Their
equations of motion are constraint equations that fix, respectively, the total energy 
and the two-dimensional curvature. More precisely, if there is a two-dimensional potential 
energy density $V(A,\tau)$ coming from the 4D cosmological constant, the Casimir energy of 
light 4D fields, and possibly other sources, then the 2D vacua are characterized 
by a vanishing potential $V = 0$ and a curvature ${\cal R}_{(2)} = \partial_A V$. On the other hand $\tau$ is dynamical, 
so in order for a 2D vacuum to be stable it should correspond to a minimum of the potential $V$ along the $\tau_{1,2}$ directions. 
We did not attempt a detailed analysis of the 2D potential in the Standard Model 
in order to find configurations $(A_0, \tau_0)$ meeting the above conditions. 

The ultimate possibility is compactifying all three spatial dimensions. 
The resulting theory is a 1D effective theory---quantum mechanics.
At low energies the degrees of freedom are the overall size of the compact manifold $a(t)$ 
and the shape moduli which we collectively denote by $\Phi(t)$. The system is described by a mechanical Lagrangian
\be
{\cal L} =   \sfrac12 M_4^2 \big[ - 6 \, \dot a^2 a+ a^3 \,
\dot \Phi \cdot K(\Phi) \dot \Phi \big] -  V(a, \Phi) \; ,
\ee
supplemented by the constraint that the total Hamiltonian vanishes, ${\cal H} = 0$, the so-called 
Hamiltonian constraint. In the Lagrangian above $K(\Phi)$ is a positive definite matrix, while $a(t)$ 
enters with a negative definite kinetic energy. Notice that in all previous cases by `vacua' we meant 
compactified solutions with maximal symmetry (de Sitter, Minkowski, or Anti-de Sitter) in lower dimensions, 
whereas here in the 1D theory all ``fields'' only depend on time, and the only sensible definition of  a vacuum 
seems to be `a time-independent solution'. However the Hamiltonian constraint makes it impossible 
for such a solution to exist, unless a perfect tuning is realized in the potential---$V$ should have 
a stationary point at which $V$ itself exactly vanishes. Indeed we are used to the fact that the Lagrangian 
above generically describes a cosmology, the Hamiltonian constraint being just the first Friedman equation.
In Appendix \ref{1Dvacua} we discuss the closest analogue we can have to a vacuum---an almost static 
micron-sized universe that undergoes classical small oscillations in size and shape on a time-scale of order of our Hubble time. 
However on longer time-scales such a system is necessarily unstable against decompactification, crunching, or asymmetric  
Kasner-like evolution, due to the wrong-sign kinetic energy of the scale factor $a$.

%
%
%
%
%
%
%
%
%
%

\section{AdS/CFT and the real world}
We have seen that, with the minimal particle content consistent with 
neutrino masses, the Standard Model has AdS$_3 \times$S$_1$ vacua, even 
though the 4D cosmological constant is small and positive. This vacuum is 
clearly a very close cousin of our own---since the size of the circle is 
$\sim 20$ microns, the high-energy physics in this vacuum---including the 
Standard Model spectrum, whatever UV completes it all the way up to the 
Planck scale, even trans-Planckian quantum gravitational physics up to 
energies up to $10^{48}$ GeV where $\sim 20$ micron black holes are 
produced---is the same as in ours.

By AdS/CFT duality \cite{AdS/CFT} there must exist a two-dimensional CFT dual to 
physics in this background. Of course this must be a very peculiar CFT. 
The central charge is $c\sim \ell_3 M_3 \sim 10^{90}$. The spectrum of 
operator dimensions is strange---there are a few operators with $O(1)$ 
dimensions, dual to the metric, the photon, the graviphoton and the 
radion. The operator dual to the Wilson line is rather bizarre---it is 
nearly marginal, with an anomalous dimension of order $e^{-10^8}$! There is  
an enormous gap till the operators dual to neutrinos and Kaluza-Klein modes on the S$_1$ are 
encountered, with dimensions of order $\sim 10^{30}$, and then even larger 
gaps to more an more irrelevant operators corresponding to the electron, 
muon, pions and the rest of the Standard Model spectrum. All the details 
of the both the Standard Model and whatever comes beyond it are contained 
in the spectrum of ridiculously irrelevant operators in the CFT.

Of course CFT's with this type of huge gap in their spectrum of operators 
have long been known to be relevant to duals of string theory models 
compactifying to AdS with fixed moduli. Indeed, the peculiarity of the 
CFT's led some to speculate that such CFT's are impossible and that there 
had to be some hidden inconsistency in these constructions. Here we see 
that precisely such CFT's arise even in the simplest possible case of 
2D theories as the duals of the AdS$_3 \times$S$_1$ vacuum of the 
Standard Model. Conversely, if it is ever possible to prove that CFT's 
with these properties do not exists, this necessarily implies that the 
deep IR spectrum of our world must have additional light states to remove 
the AdS minimum of the radion potential!

How is the $SU(3) \times SU(2) \times U(1)$ gauge symmetry of the Standard Model reflected in the CFT? Ordinarily we associate gauge symmetries
in the bulk with global symmetries on the boundary; there is clearly a global $U(1)$ current associated with the long-distance 
bulk $U(1)_{EM}$ gauge symmetry, but what about the gauge symmetries that have been Higgsed and confined at energies far above the AdS curvature 
scale? These are not simply reflected in the CFT, which is appropriate--there are no massless degrees of freedom associated with them in the bulk, 
and as gauge symmetries are just redundancies of description, 
the CFT should only contain the gauge-invariant physical information---such as the spectrum of hadrons and the electroweak 
symmetry breaking masses.

Our AdS$_3$ minima are certainly metastable; there may be deeper AdS$_3$ minima in the Standard Model landscape. Whatever the deepest such 
minimum is, could it absolutely stable? This would be surprising given that the background is completely non-supersymmetric. One possible 
instability would be the nucleation of a Witten bubble of nothing \cite{nothingbubble} but this requires antiperiodic fermions around the circle while our vacuum
exists only for periodic fermions. A more fundamental issue is that, since the bulk 4D theory has a positive cosmological constant and a 
dS$_4$ vacuum, we expect on general grounds that this dS$_4$ solution should be unstable to tunneling into other parts of the larger 
landscape. The dS$_4$ decays via bubble nucleation; the bubble size $R_4$ can range in size from micro-physical scales to as large as the 
dS$_4$ Hubble length $\ell_4$, the latter arising from the minimal possibility of Hawking-Moss transitions out of de Sitter space on 
Poincare recurrence times.    
 If our cosmological constant is tuned to be be small by the presence of a huge discretum of nearby vacua, $R_4$ is parametrically smaller than $\ell_4$,
 it is conceivable that $R_4 \sim \ell_4$ if our vacuum is isolated by huge potential barriers from the rest of the landscape. 
 
 How is the apparently necessary dS$_4$ instability reflected in the CFT$_2$ dual of the AdS$_3$ vacuum? 
 If $R_4$ is smaller than the size of the S$_1$, $\sim 20$ microns, it is 
clear that the same bubble nucleation process will occur in the AdS$_3 \times$S$_1$ vacuum. 
Actually, even if $R_4$ is only smaller than the 
AdS$_3$ length, there is an effective 3D bubble that mediates the decay: if the domain wall bounding the surface of the 4D bubble 
has surface tension 
$\sigma$ and the energy difference between vacua is $p$, we have $R_{4} \sim \sigma/p$; 
wrapping the wall on the circle gives us a lower-dimensional wall of tension $\sigma r$ while the pressure difference is $p r$, so 
a 3D bubble 
has a size $R_3 \sim (\sigma r)/(p r) \sim R_{4}$. For our neutrino-supported AdS$_3$ 
vacua, the AdS$_3$ length is only few times smaller than the dS$_4$ Hubble; so 
if there is a discretum of vacua allowing for the adjustment of our cosmological constant, 
the AdS$_3$ must also 
be unstable, with an exponentially long lifetime.  

Presumably this means that the CFT must itself be ill-defined at a tiny non-perturbative level~\footnote{We thank Juan Maldacena and Nathan Seiberg for 
a discussion on this point.}; for instance by having a marginal 
perturbation $g$ with a metastable minimum and an unbounded below potential. The timescale of the instability of the CFT could be
of order $\ell_3\, e^{-1/g} $. If our vacuum is isolated by huge barriers from the rest of the landscape, it is conceivable 
that the AdS vacua are absolutely stable, since the required bubble, while being smaller than $\ell_4$, could be larger than $\ell_3 \sim \ell_4/4$, 
though this seems incredibly unlikely!

%
%
%
%
%
%
%
%
%
%

\section{Quantum Horizons}
\label{quantumhorizons}
\begin{figure}[t]
\begin{center}
\begin{picture}(550,150)(0,35)
\put(40,20){\epsfig{file=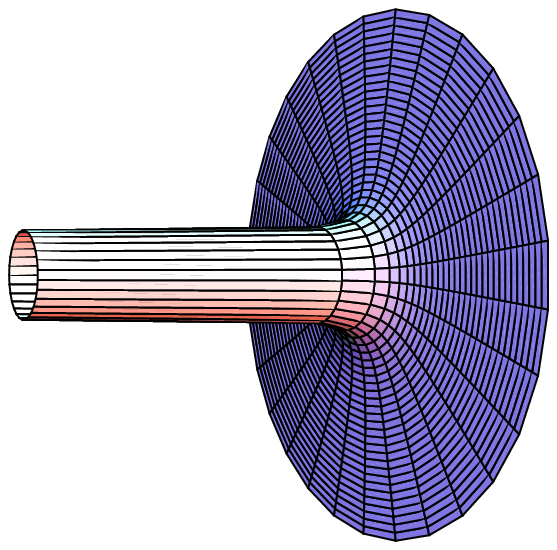,width=5cm,height=6cm}}
\put(120,20){a)}
\put(230,30){\epsfig{file=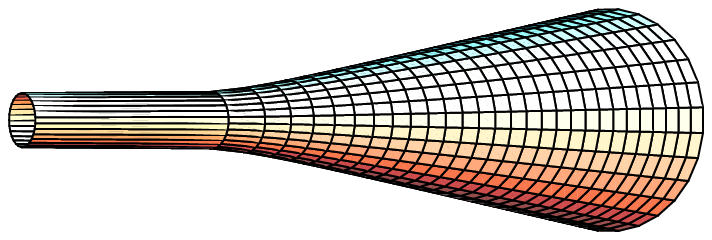,width=8cm,height=5cm}}
\put(370,20){b)}
\end{picture}
\end{center}
\caption{\small\it {\rm a)} Extremal Reissner--Nordstrom black hole  interpolates between asymptotically flat region and AdS$\times$S vacuum; 
{\rm b)} Extremal black hole interpolating to the lower dimensional vacuum stabilized by the Casimir effect.}
\label{throat}
\end{figure}

Given the existence of a landscape of vacua in the Standard Model, it is natural to ask whether it is possible to find
geometries interpolating between vacua with a different number of non-compact space dimensions. 
Such interpolations are already familiar for classical AdS$_n\times$S$_m$ vacua. 
For instance, the Standard Model possesses  AdS$_2\times$S$_2$ vacua
with the sphere stabilized by a flux of the electric field. The interpolating solution is nothing but the extremal
Reissner--Nordstrom black hole. Indeed, far from the black hole the metric is flat, while in the vicinity of the  horizon 
an infinite AdS$_2\times$S$_2$ throat  is developed, see Fig.~\ref{throat}a. 

It has been fruitful to view extremal black holes as interpolations between different vacua  (cf. \cite{Gibbons:1993sv}) in the context of
the interpolation of the scalar moduli fields in supersymmetric theories between spatial infinity and the black hole horizon (``attractor mechanism"). As we will show, 
this viewpoint is useful in broader context. In particular, for  the Casimir stabilized vacua this leads to black hole solutions with the horizon 
supported entirely by the quantum effects (the Casimir energy).

There is a qualitative difference between classical AdS$_2\times$S$_2$ and the Casimir vacua. In the former case the radius of the sphere $S_2$ is
of the same order as the AdS$_2$ length, while for the Casimir vacuum we have a true compactification, where the size of the compact space  is much smaller than
the curvature length along the non-compact coordinates. Related to this, in the Casimir  compactification the radion mass $m_R$ is much lighter than 
the compactification scale $1/R$. The length scale at which the interpolation happens is determined by the inverse mass of the radion $1/m_R$. This is
of order $R$ for the classical AdS$_2\times$S$_2$ vacuum so the interpolating geometry has a form of a ``hole".  Instead, for the Casimir vacuum the interpolating
geometry takes the form of a cone with a narrow opening angle (see Fig.~\ref{throat}b)).

\label{Adsdinterpol}
\subsection{Setting up the problem in the three-dimensional case}
\label{setting}
Let us start a more explicit analysis by exploring geometries interpolating between three-dimensional and two-dimensional
vacua. As we will see this case turns out to be remarkably simple technically
but contains much of non-trivial physics. It is natural to look for an interpolating metric with the following form
\be
\label{3Dinter}
ds^2=-A^2(z)dt^2+dz^2+R^2(z)d\phi^2\,,
\ee
where $\phi\in[0,2\pi)$ is a periodic coordinate.
The precise form of the energy-momentum tensor is determined by the specific mechanism used to stabilize the two-dimensional
vacuum. It is straightforward to calculate  the energy-momentum related to the classical contributions to the radion
potential. 
For instance, if the cosmological constant in three dimensions is negative, one can obtain a stabilized 
lower dimensional AdS$_2\times$S$_1$ vacuum by turning on a flux of a scalar axion field $\Phi$ (note, that in three
dimensions this is equivalent to having the electromagnetic flux),
\[
\Phi=F\phi\;.
\]
Then, by explicit computation, the axion energy-momentum tensor in the geometry (\ref{3Dinter}) is 
\be
\label{TMN}
T_{M}^{N}=-
\l
\begin{array}{cc}
\rho(R)\delta_\mu^{\nu} & 0\\ 
0&\sigma(R) 
\end{array}
\r\;,
\ee
where 
\[
\rho(R)=\frac{F^2}{R^2}
\] 
is the classical contribution to the radion potential coming from the gradient energy of the axion field,
and
\be
\label{Tffrel}
\sigma(R)=\rho(R)+R\,\d_R\rho(R)\;.
\ee
Actually, the relation (\ref{Tffrel}) between $T_\phi^\phi$ and $T_\mu^\nu$ is a direct consequence of the
conservation of the energy-momentum tensor of the form (\ref{TMN}) in the metric (\ref{3Dinter}).

This was a classical example. The Casimir contribution
to the energy-momentum for the geometry (\ref{3Dinter}) is in principle more involved. Indeed,  the compactification scale $R$ is changing
in space,  so the one-loop contribution is a complicated functional depending on  the local value of $R(z)$ as 
well as on all its derivatives. Fortunately, we do not need the exact form of this functional for our purpose of finding the interpolating
solutions. Indeed, as we argued before, we expect $R(z)$ to be a very slow varying function of $z$, so that locally the geometry is well
approximated by a cylinder, and the derivative part of the Casimir energy can be safely ignored. 
Under this assumption, because of the Lorentz invariance, the energy-momentum tensor is again of the form (\ref{TMN})
where $\rho(R)$ is  determined by the Casimir energy,
\[
\rho(R)=\frac{V_C(R)}{2\pi R}\;.
\]
We proceed with general $\rho(R)$, and will be more specific about its shape later, when necessary.
Of course,  as shown in the Appendix~\ref{app:Casimir}, a $T^N_M$ of the form (\ref{TMN}) agrees with 
the explicit calculation of the Casimir
energy-momentum. 
 
To summarize, we need to  study  solutions of the three-dimensional Einstein equations for the metric
ansatz (\ref{3Dinter}) with the energy-momentum of the form (\ref{TMN}). Explicitly, these equations are
\begin{eqnarray}
\label{tt}
M_3\, R''&=&-R\,\rho(R)\;,\\
\label{zz}
M_3\, A'\, R'&=&-A\,R\,\rho(R)\;,\\
\label{ff}
M_3\, A''&=&-A\left [ \rho(R)+R\,\d_R\rho(R) \right]\;,
\end{eqnarray}
where $M_3$ is the three-dimensional Planck mass.
For  the  two-dimensional vacua the radius of the compact dimension is constant $R=R_0$
so they correspond to zeros of the Casimir energy,
\[
\rho(R_0)=0\;,
\]
while the  curvature along the non-compact dimensions is determined by the slope of $\rho$,
\be
\label{vacua}
A(z)=\left\{
\begin{array}{ccc}
\exp(z/\ell_2),&\;{\rm AdS}_2\times {\rm S}_1\;\mbox{vacuum}\; &(\rho'<0)\\
1,&\;{\rm M}_2\times {\rm S}_1\;\mbox{vacuum}\;&(\rho'=0)\\
\cos(z/\ell_2),&\; {\rm dS}_2\times {\rm S}_1\mbox{vacuum}\;&(\rho'>0)
\end{array}
\right. \,.
\ee
These coordinates cover the Poincare and causal patches of AdS$_2$ and dS$_2$, but can be clearly extended
to the global AdS$_2$ (dS$_2$).

For solutions with non-constant $R(z)$ one can take the ratio of the $(tt)$ and $(zz)$ equations (\ref{tt}) and (\ref{zz}),
and arrive at the following relation between $A$ and $R$,
\be
\label{ttzz}
A(z)=R'(z)\;.
\ee
As a result the interpolating metric (\ref{3Dinter}) takes the form
\be
\label{3Dsimple}
ds^2=-R'^2 dt^2+dz^2+R^2 d\phi^2
\ee
and the $(tt)$ equation (\ref{tt})
implies that $R$ is a solution to the one-dimensional mechanical problem
with the effective potential $U$  determined by
\be
\label{Ueff}
\frac{dU}{dR}=M_3^{-1} R\,\rho(R)\;.
\ee
Note that this potential is extremal at the values of $R$ corresponding to the lower
dimensional vacua.
From eq.~(\ref{3Dsimple}) 
we see a direct confirmation of the intuition that the interpolation to the
lower dimensional vacuum takes place in the near horizon limit---the region 
where the radius of the compact dimension approaches a constant value, $R'\to 0$, corresponds to the
horizon of the metric (\ref{3Dsimple}). 
To understand better the causal structure of the metric
(\ref{3Dsimple})  
it is convenient to perform a change of coordinates and use $R$ itself as the interpolating variable.
With this choice of coordinates the metric (\ref{3Dsimple}) is
\be
\label{3DU}
ds^2=-f(R)dt^2+f(R)^{-1}dR^2+R^2 d\phi^2\;,
\ee
where $f(R)=R'^2$ can be found explicitly by making use of the ``energy" conservation law of the mechanical problem (\ref{tt}), giving
\be
\label{energyconservation}
f(R)\equiv R'^2=\epsilon-U(R)\;.
\ee
The $(tR)$ part of the metric (\ref{3DU}) has the typical form of black hole geometries, with
horizons located where $f(R)$ is zero. We see that the metric ansatz (\ref{3Dinter}), having the advantage
of making the interpolating nature of the solution explicit, actually covers only a small part of the interpolating geometry.

For metric  written in the form (\ref{3Dinter}) we found two branches of solutions---lower dimensional
vacua (\ref{vacua}) and solutions with non-constant $R(z)$, described by the mechanical problem
(\ref{Ueff}). The latter  can be presented in the form (\ref{3DU}). To recover the compactified vacuum
solutions with (\ref{3DU}) let us choose $\epsilon=U(R_0)$ 
and zoom on the part of the geometry (\ref{3DU}) where $R$ is close to $R_0$. 
Namely, let us write
\[
R=R_0(1+\alpha\, r)
\]
and rescale $t\to \tau=\alpha\, R_0\, t$. Taking the limit $\alpha\to 0$ we obtain the AdS$_2\times$S$_1$ 
(dS$_2\times$S$_1$) metric  for negative (positive) $U''(R_0)$ in the form (\ref{3DU}),
\[
ds^2=-\frac{U''(R_0)}{2}r^2 d\tau^2+\frac{2}{U''(R_0)} \frac{dr^2}{r^2}+R_0^2 d\phi^2\;,
\]
 with curvature length
\be
\label{AdSlength}
\ell_2=\sqrt{\frac{2}{|U''(R_0)|}}\;.
\ee

Finally, let us recall that our solutions are trustworthy as long as the
radius $R$ changes slowly along the non-compact coordinates. When the metric is written in the form 
(\ref{3Dsimple}) this implies $R'\ll 1$. From the definition (\ref{energyconservation})
we see that this condition is translated in the frame (\ref{3DU}) to
\[
f(R)\ll1\;.
\]
Given that this condition is satisfied one can trust the metric (\ref{3DU}) both in the regions where
$f(R)$ is positive and negative, {\it i.e.} irrespectively of whether the size of the compact dimension
changes in space or time.

\subsection{Extremal black holes interpolating from M$_3$ to AdS$_2\times$S$_1$}
\begin{figure}
\psfrag{R}{\footnotesize $R$} \psfrag{UR}{\hspace{-14pt} \footnotesize $U(R)$}
\psfrag{I1}{\tiny \bf \textcolor{blue}{I}}\psfrag{I2}{\tiny \bf \textcolor{blue}{II}}\psfrag{I3}{\tiny \bf \textcolor{blue}{III}}
\psfrag{Rinf}{\hspace{-2pt}\tiny \textcolor{red}{$R$=$\infty$}}\psfrag{R0}{\hspace{-7pt}\tiny \textcolor{red}{$R$=$0$}}
\psfrag{singularity}{\hspace{-20pt} \footnotesize naked singularity}
\psfrag{hor}{\hspace{-4pt}\footnotesize horizon}\psfrag{atinf}{\footnotesize at $\infty$}
\begin{center}
\vspace{-.6cm}
\begin{tabular}{|c|c|c|}
\hline
\raisebox{1cm}[0pt][0pt]{\epsfig{file=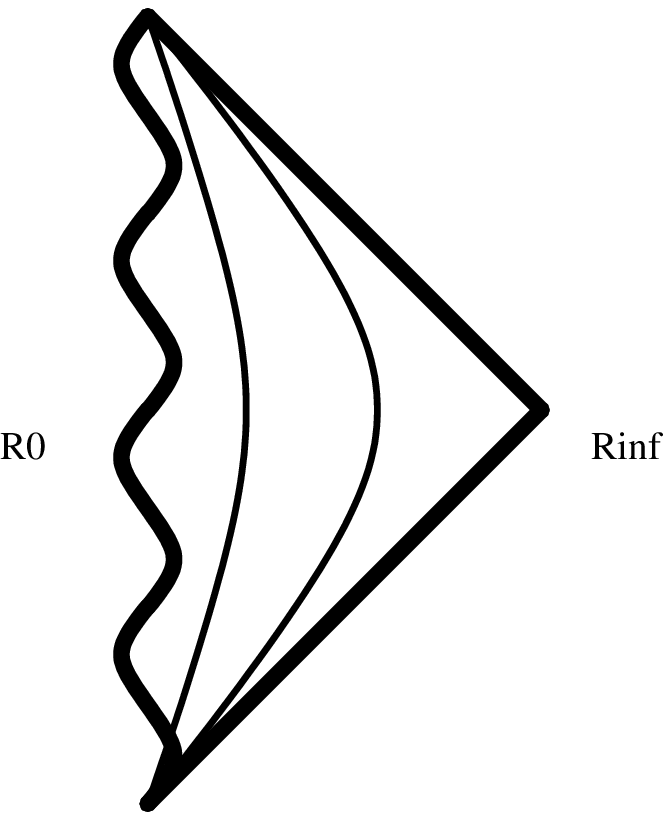,height=2cm}}&
\epsfig{file=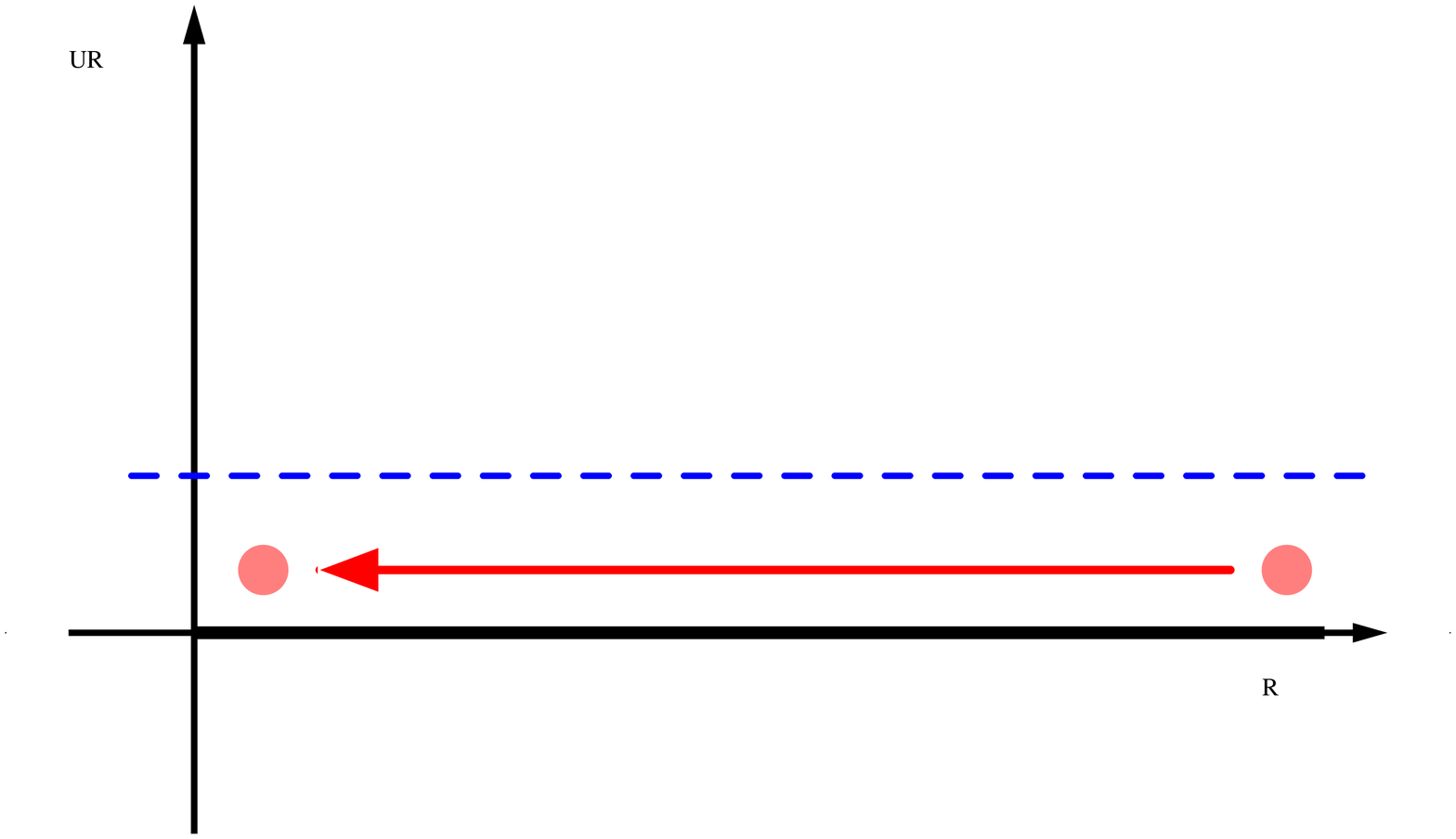,height=4cm} & 
\raisebox{1cm}[0pt][0pt]{\epsfig{file=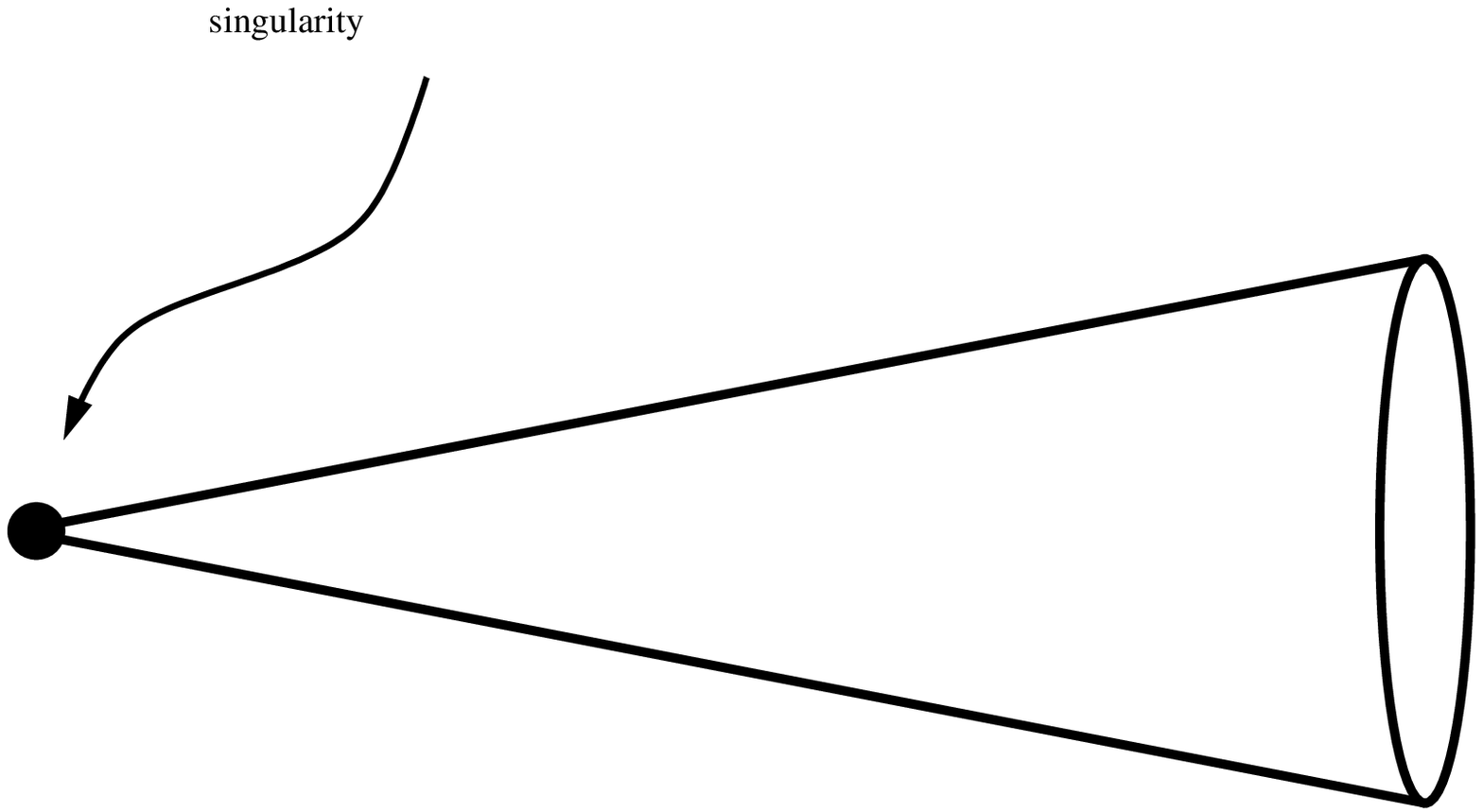,height=2cm}}\\ 
\hline
\hline
\hline 
\raisebox{1cm}[0pt][0pt]{\epsfig{file=Pen-super.eps,height=2cm}} &
\epsfig{file=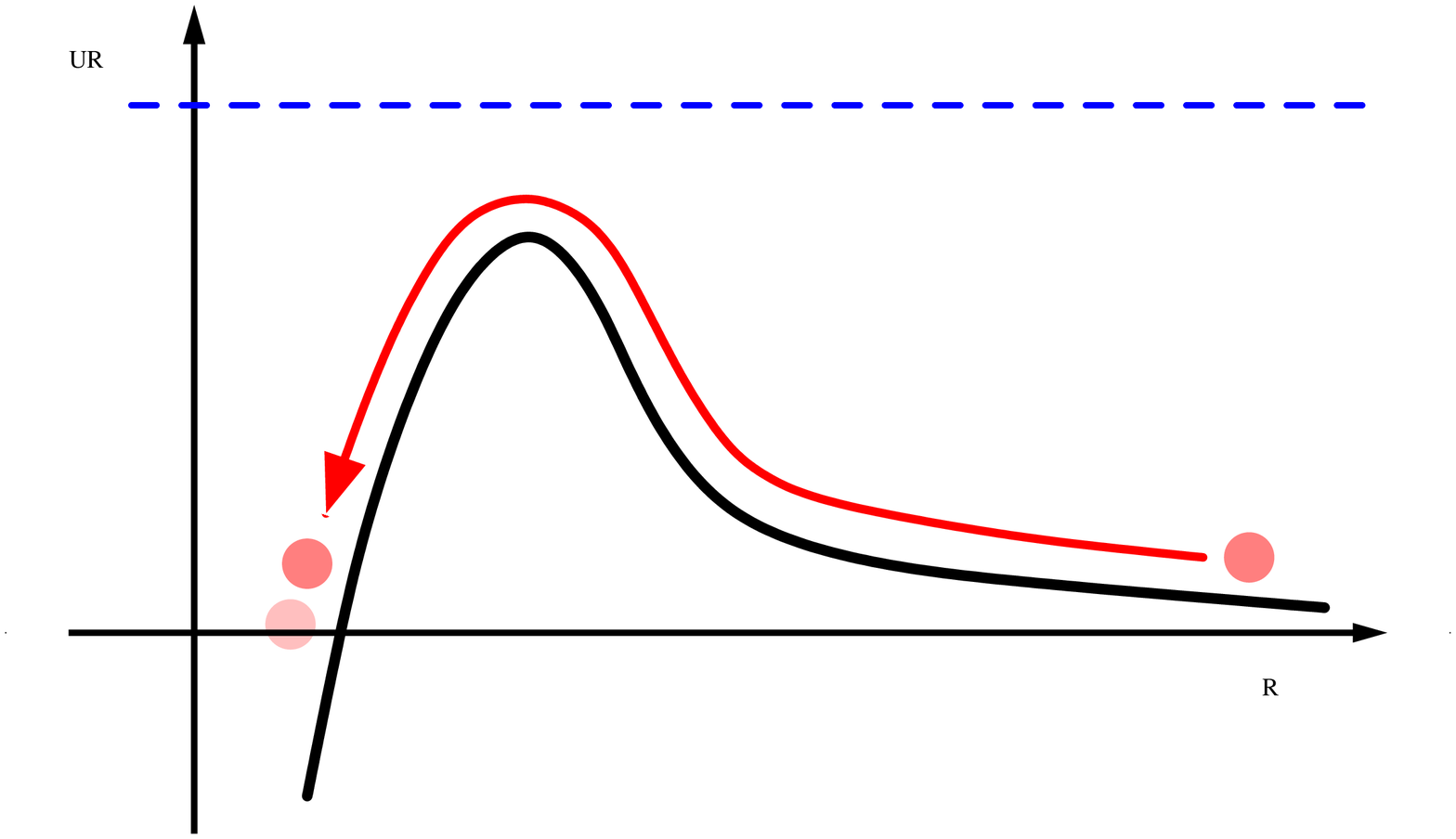,height=4cm} & 
\raisebox{1cm}[0pt][0pt]{\epsfig{file=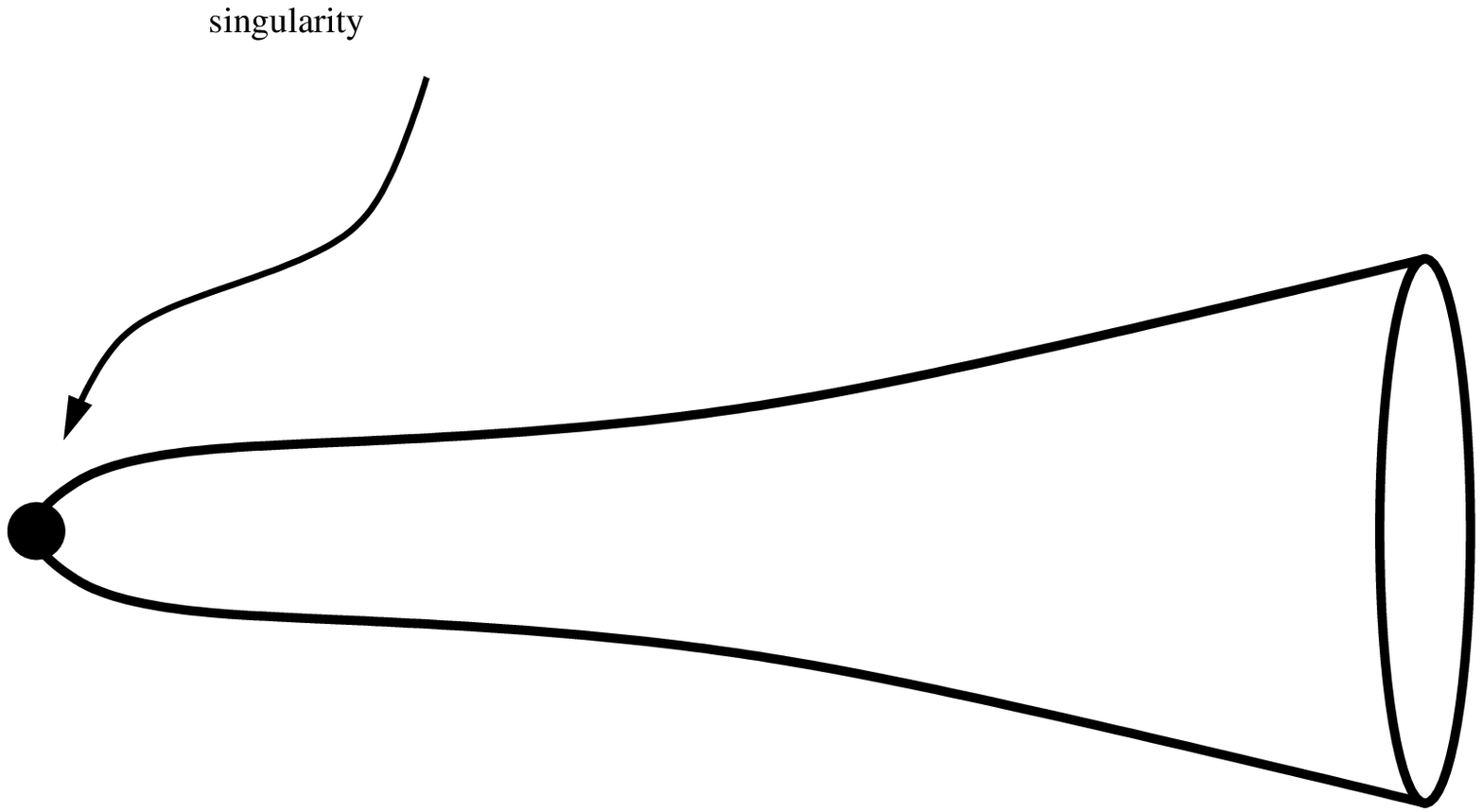,height=2.5cm}}\\ 
\hline 
\raisebox{0.2cm}[0pt][0pt]{\epsfig{file=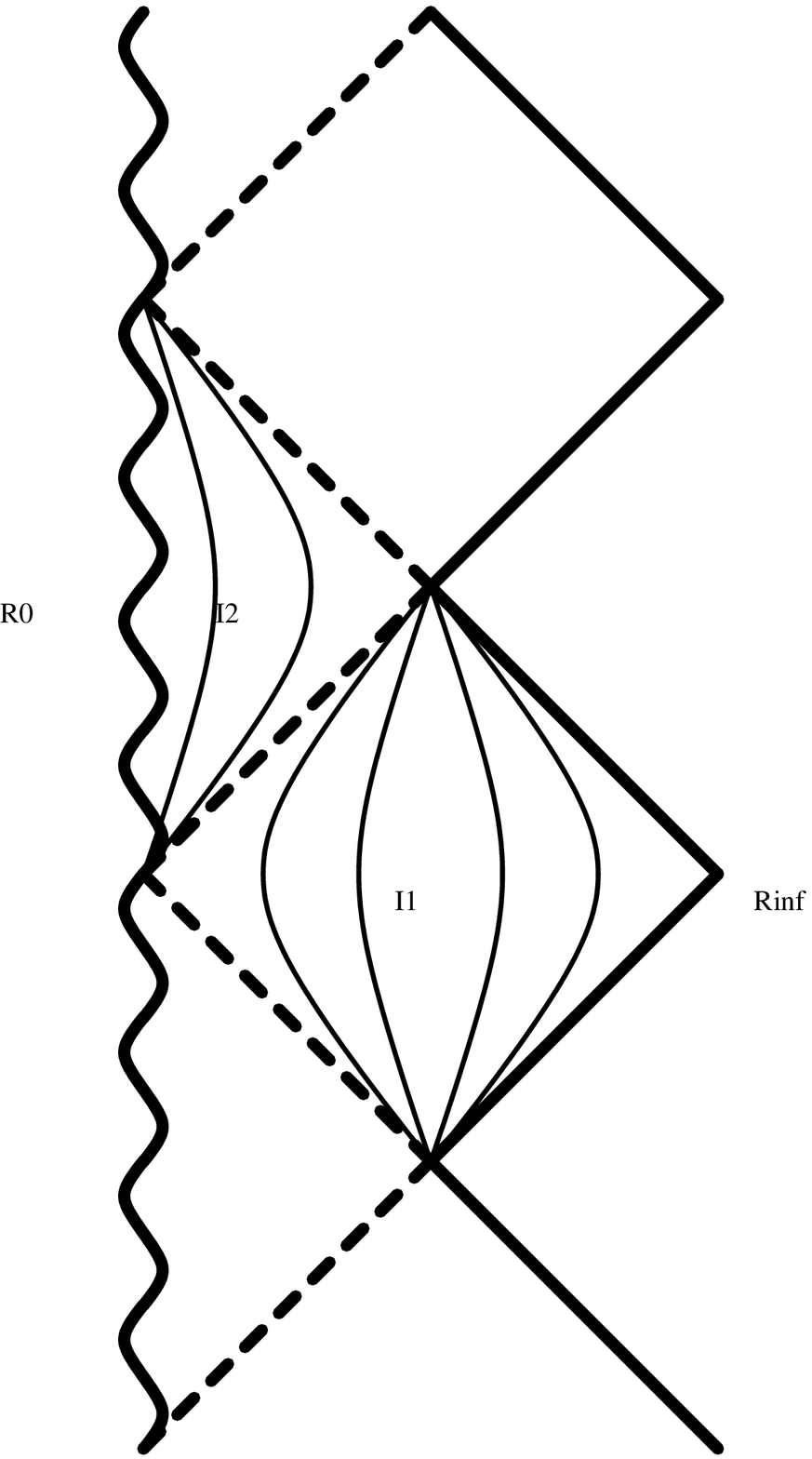,height=3.5cm}}&
\epsfig{file=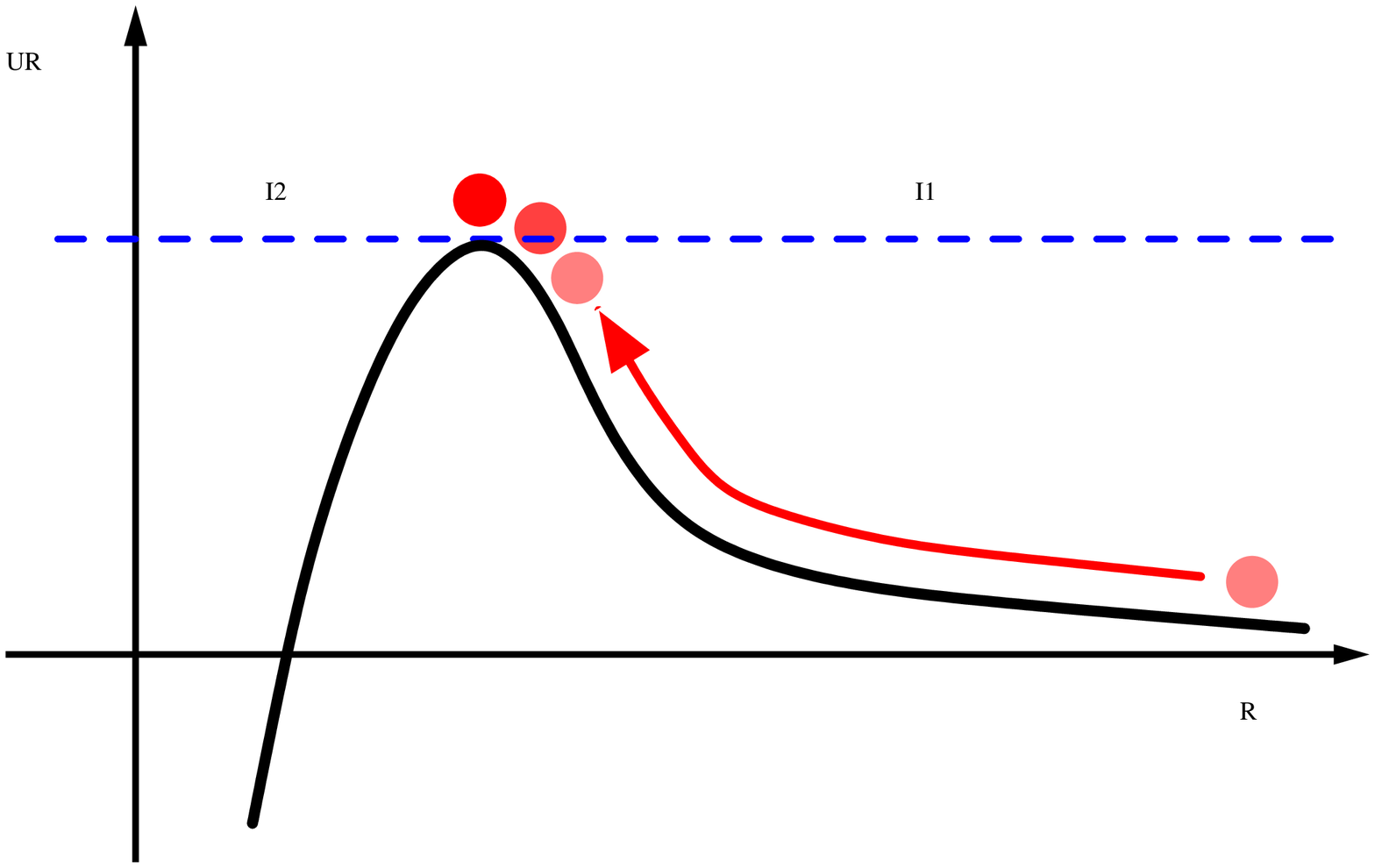,height=4cm} & 
\raisebox{1cm}[0pt][0pt]{\epsfig{file=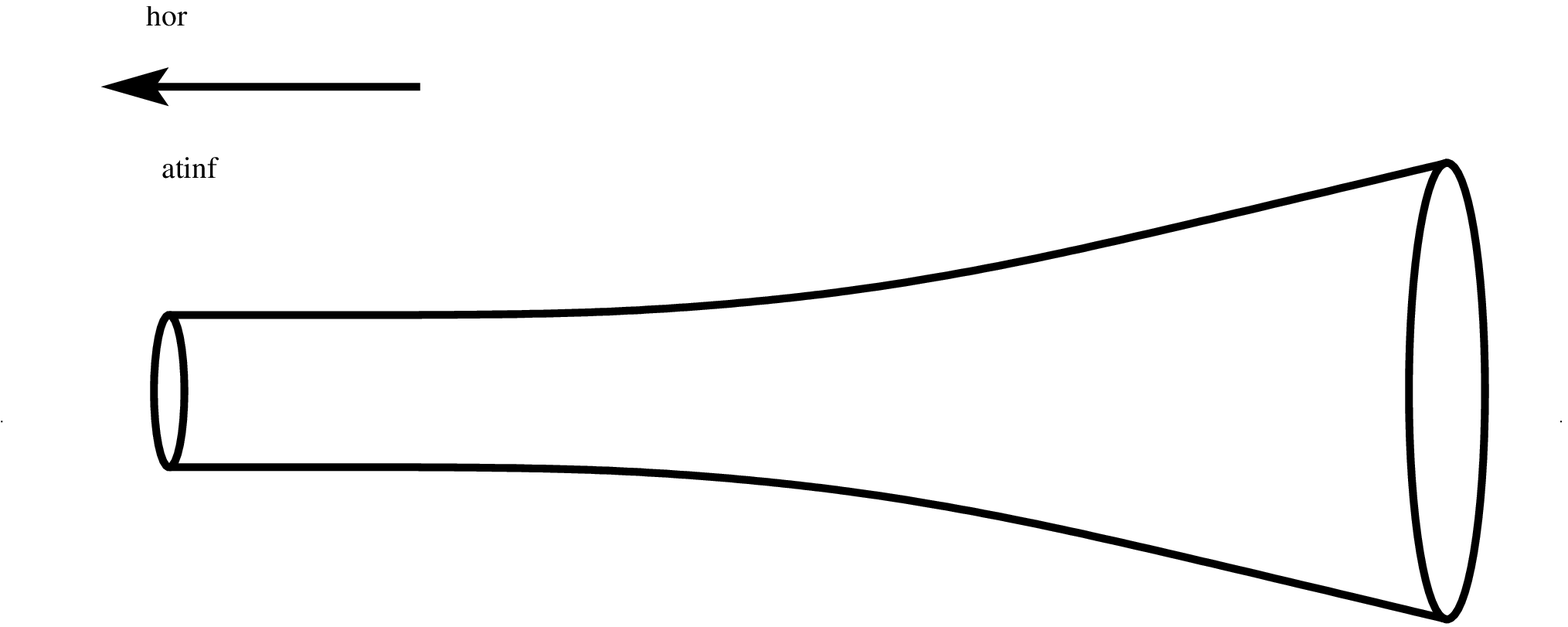,height=2.2cm}}\\ 
\hline  
\raisebox{0.2cm}[0pt][0pt]{\epsfig{file=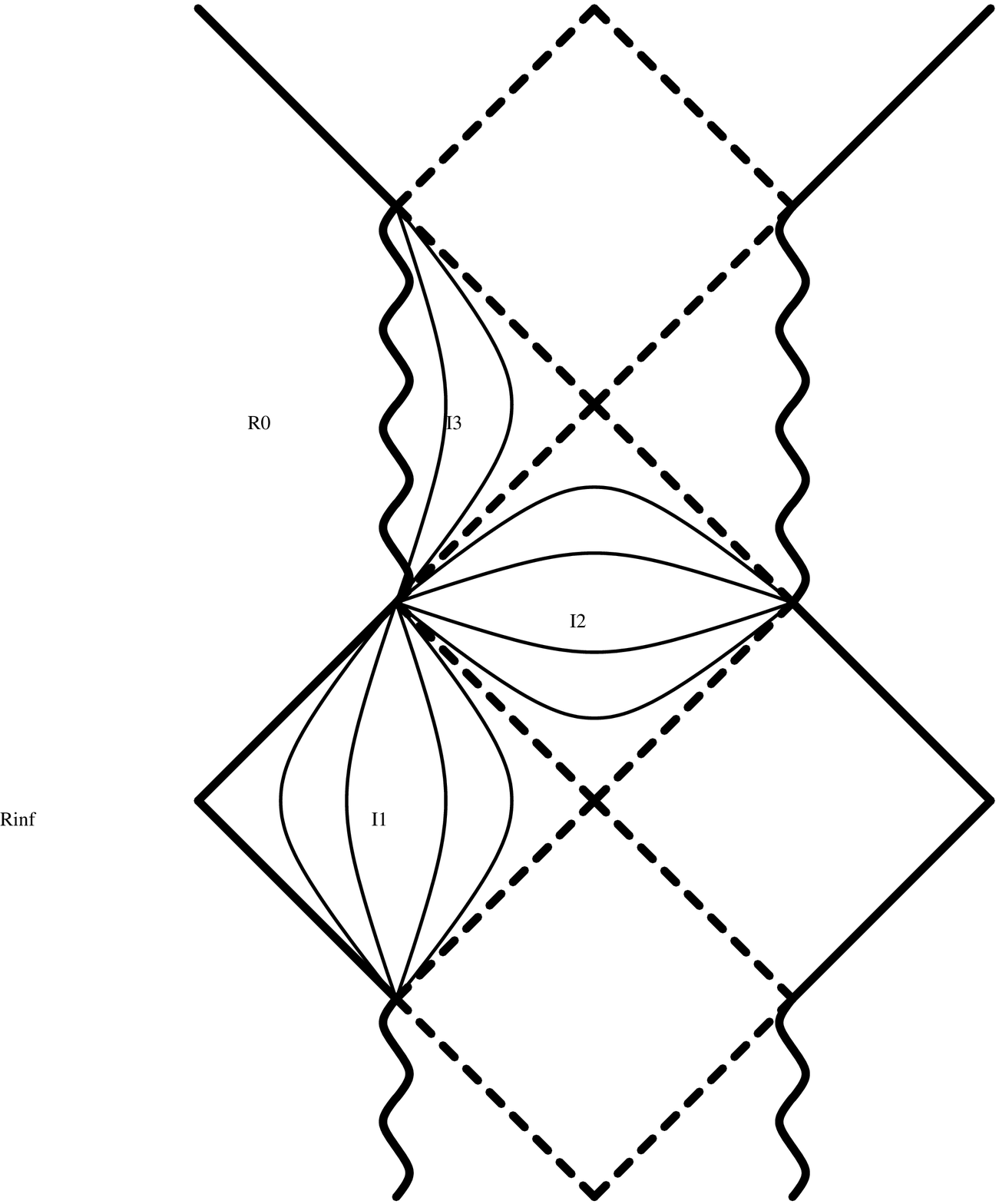,height=3.5cm}}&
\epsfig{file=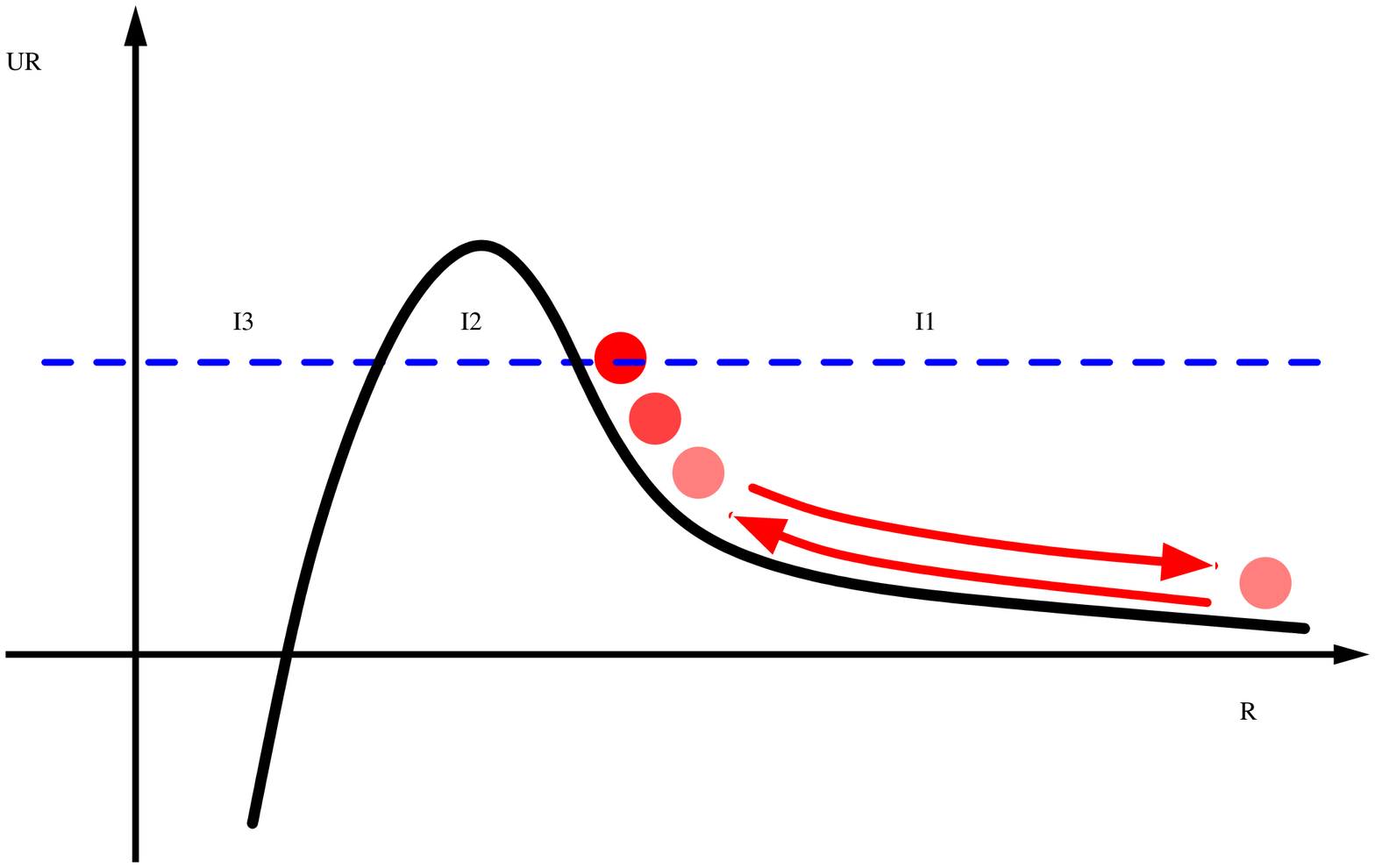,height=4cm} & 
\raisebox{1cm}[0pt][0pt]{\epsfig{file=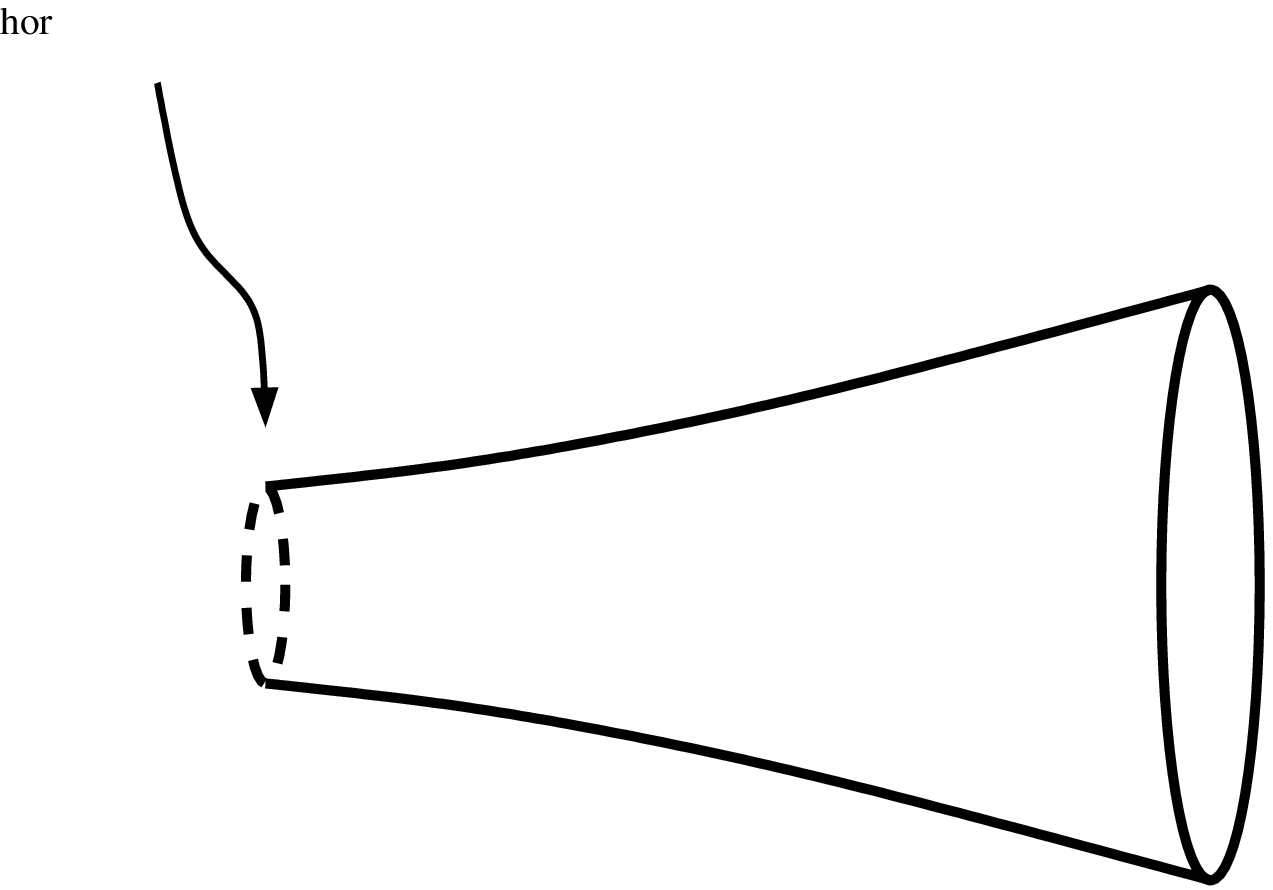,height=2.5cm}}\\ 
\hline  
\raisebox{0.4cm}[0pt][0pt]{\epsfig{file=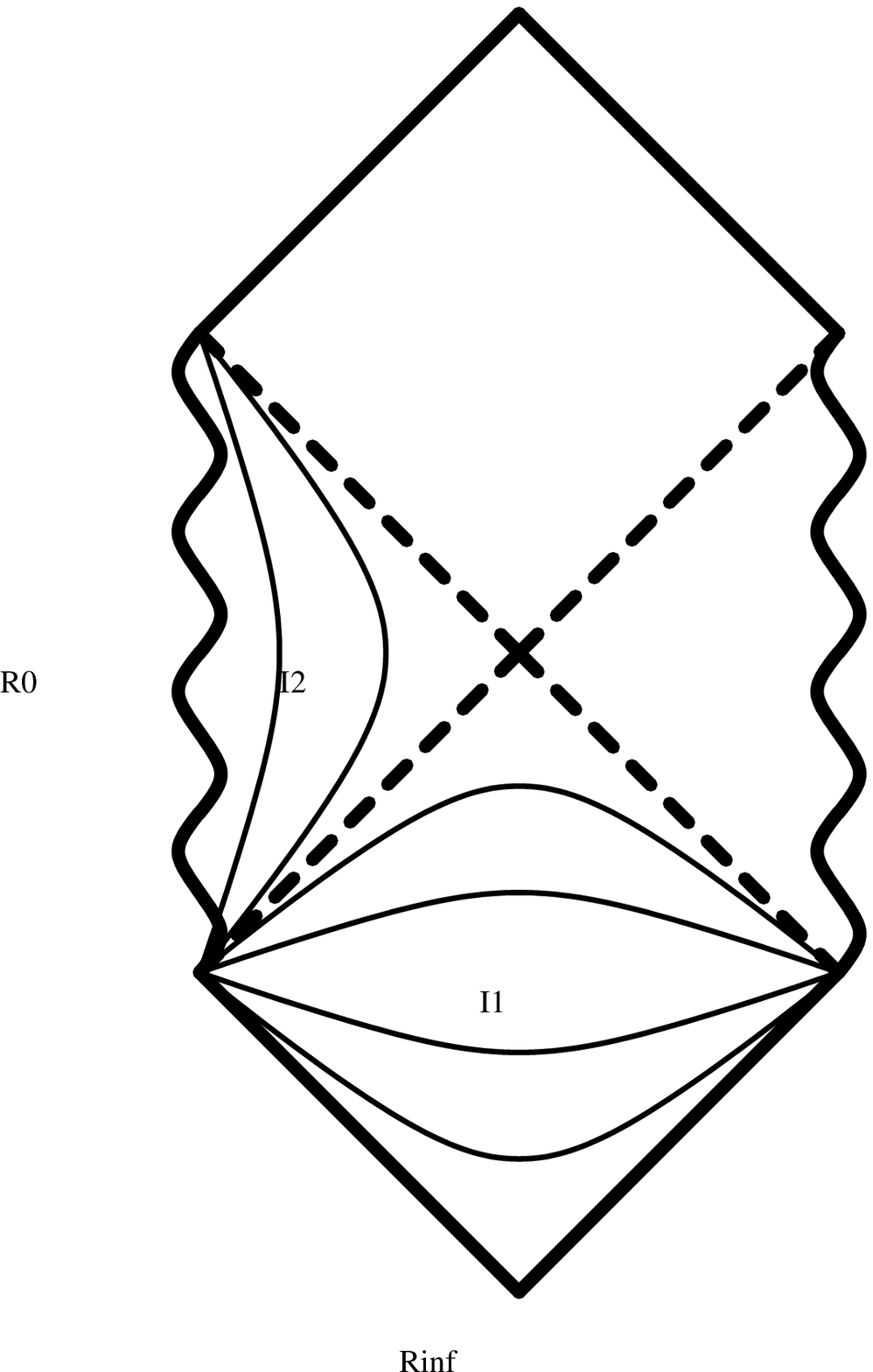,height=3cm}}&
\epsfig{file=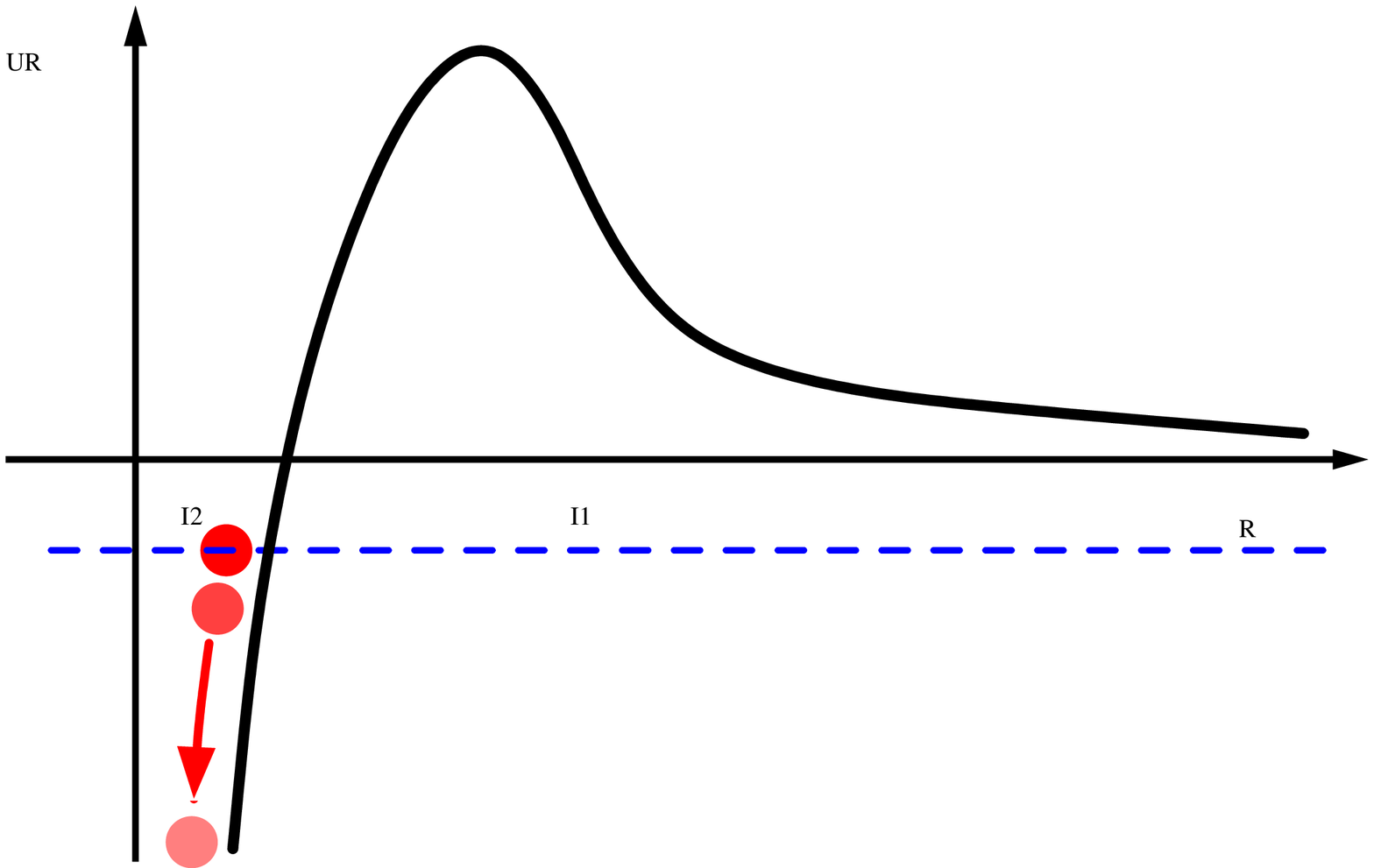,height=4cm} & 
\raisebox{1.4cm}[0pt][0pt]{\epsfig{file=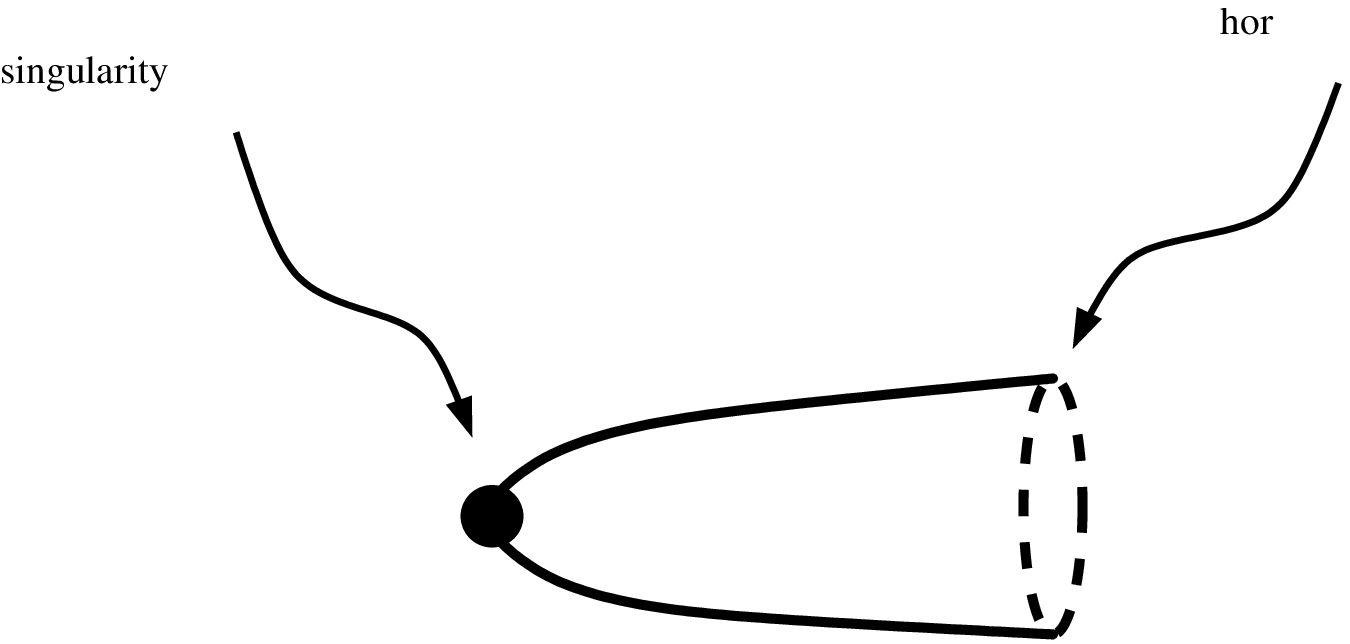,height=2cm}}\\ 
\hline 
\end{tabular}
\end{center}
\caption{{\small \it Interpolating solutions from M$_3$ to AdS$_2$ for 
trivial (1st row) and non-trivial Casimir (2nd$\div$5th row) 
mechanical potential (drawn in the 2nd column) $U(R)$, defined in eq.~(\ref{Ueff}) 
for different choices of the total energy (blue dashed line).
In the first column there are the corresponding Penrose diagrams for the extended solution (eq.~(\ref{3DU})),
with iso-$R$ curves explicit. 
In the third column the actual geometry is shown for the classical allowed 
regions of the potential (corresponding to the red arrow curve).
}}
\label{fig:ads3D}
\end{figure}

To be concrete, let us first focus on the case when the three-dimensional cosmological constant is zero and the axion fluxes are absent,
so that the effective potential $U$ goes to zero at large values of $R$. Let us start with the simplest case,
when  the lower dimensional vacuum has  a negative cosmological constant. According to (\ref{vacua}) this 
implies that the effective potential has a maximum at $R=R_0$, see Fig.~\ref{fig:ads3D}. 
A potential of this form 
is generated, for instance,  in a theory with some number of light  bosons and   
heavy fermions, such that the total number of fermionic degrees of freedom is larger than the total number of bosons. 
In the simple case when all fermion masses  are characterized by a single scale $\mu\ll M_3$    all other parameters in the effective
potential $U$ are also determined by this scale. For instance, the radius of the compact dimension
in the two dimensional vacuum 
is 
\[
R_0\sim \mu^{-1},
\]
while the value of the potential at the maximum is
\[
U(R_0)\sim \frac{\mu}{M_3}\;.
\]
\pagestyle{empty}
\thispagestyle{plain}
The solution to the one-dimensional  mechanical problem, which determines the shape $R(z)$ of the interpolating
geometry, has energy $\epsilon=U(R_0)$, so that it reaches the top of the effective potential in an infinite ``time" $z$.
For $z\to +\infty$ it describes a flat ($2+1$) dimensional space
with opening angle equal to
\be
\label{eps}
\theta_o=2\pi\left.R'\right|_{z=+\infty}=2\pi\sqrt{U(R_0})\sim 2\pi\l\frac{\mu}{M_3}\r^{1/2}\;.
\ee
At $z=-\infty$ the radius of the compact dimension is exponentially approaching its stabilized value
\[
R=R_0(1+e^{-|z|/\ell_2})\;,
\]
so that the metric (\ref{3Dsimple}) indeed asymptotes to AdS$_2\times$S$_1$ where the curvature radius
$\ell_2$ is determined by (\ref{AdSlength})
\[
\ell_2\sim  \mu\l\frac{M_3}{\mu }\r^{1/2}\;.
\]
So, as expected, the interpolating geometry has the form of a narrow cone 
with a conical singularity resolved into an
infinitely long tube.
The circumference of the horizon of the interpolating black hole is
\be
\label{L}
L=2\pi R_0\sim 2\pi \mu^{-1}\;.
\ee
 The surface gravity 
at the horizon,  which is proportional to the first derivative of the function
$f(R)$,  vanishes, so that the Hawking temperature is zero. So, we indeed obtained the
 asymptotically flat  extremal black hole solution in three dimensions. 
These are not possible in classical gravity, but accounting for the Casimir
effect leads to the appearance of the quantum horizon.  It is worth stressing again, 
that the existence and the shape of these quantum black holes is under full control in the limit
when the opening angle is small, which is  true whenever the fermion mass scale is parametrically 
smaller than the Planck mass. 

Interestingly, the Bekenstein entropy for these solutions is determined just by the  classical
geometry (opening angle) and 
does not depend on the Planck mass,
\be
\label{Bek}
S=\frac{L}{4 G_3}\propto\frac{1}{\theta_o^2}\,,
\ee
where $G_3=1/(8\pi M_3)$.
In particular, the entropy remains finite in the decoupling limit, when one sends $M_3$ to infinity while keeping
an opening angle fixed. This is understandable, 
because the mere existence of the three dimensional black holes is due to the quantum effects, so their number of microstates should remain
finite in the limit $\hslash\to 0$. 
In this limit the quantum horizon  shrinks to zero, so that one is left with a non-gravitational theory
on a cone.
Interestingly, this is similar to what happens to the extremal supersymmetric black holes in string theory, where
the Bekenstein entropy also remains finite in the limit of zero string coupling. This is one of the crucial ingredients
allowing to perform the microscopic calculations of  the black hole entropy by counting the 
BPS D-brane configurations in the decoupling limit \cite{Strominger:1996sh}. It would be very interesting to understand what are the
relevant microscopic degrees of freedom in the decoupling limit for the string realization of our 
(non-supersymmetric!) setup. 

If this is an extremal black hole what charge does it carry? Recall that the low dimensional vacuum
only exists with {\it periodic} boundary conditions for fermions. This is an ``exotic" choice. 
In general, on any simply connected
space that asymptotes to a cone, fermions would be antiperiodic in the conical region 
(for instance, if we replaced the black hole with a smooth ``cigar" tip).
This antiperiodicity is a reflection of the  ``minus" sign that the fermionic wave function 
picks up if one performs a $2\pi$ rotation around the tip.
Choosing the periodic boundary condition on the semi-infinite cylinder corresponds to switching on the 
${\mathbb Z}_2$ flux of the spin connection at the tip, similarly to how the non-integer Aharonov--Bohm flux changes  
the periodicity of the fermion wave function around a solenoid. This is the flux that labels our interpolating solution.
\thispagestyle{plain}
\pagestyle{plain}

 \subsection{Non-extremal quantum black holes}
 \label{nonextremal}
The above discussion makes it natural to look for a family of non-extremal quantum black holes carrying
${\mathbb Z}_2$ flux, such
that the interpolating solution is the limiting point for this family with the minimum mass. Also one may wonder 
whether quantum black holes exist in the sector with trivial flux (anti-periodic conditions for  fermions).
It is straightforward to identify what are these non-extremal black holes. Let us start with the charged ones and  look at the solutions to our mechanical
problem with different values of  the energy $\epsilon$. If $\epsilon>U(R_0)$, i.e., the conical opening angle at 
$z=+\infty$ is larger than for the extremal solution, a solution in the analogue mechanical 
problem overshoots $R=R_0$ and the function $f(R)$ does not have zeroes. This means that the Casimir energy is not strong enough
to shield the tip of the cone by the horizon, and a naked conical singularity develops.

 On the other hand, for values of $\epsilon$ smaller than the energy at  the top of the effective potential $U$, 
the solution to the analogue problem undershoots $R_0$. As a result $f(R)$ is zero at the turning point $R_h>R_0$,
implying that the conical singularity is shielded by a horizon. There is also an inner horizon corresponding to the 
second zero of $f(R)$, so the causal structure of the extended solution is similar to that of the 
conventional Reissner--Nordstrom black hole.

 Unlike the extremal ones these black holes have non-zero Hawking temperature.
It is most easily found by performing
the Wick rotation and identifying the periodicity
of the Euclidean time. As usual, one obtains that the Hawking temperature  is determined by the surface gravity, or explicitly,
\be
\label{TH}
T_H=\frac{f'(R_0)}{4\pi\epsilon^{1/2}}\;.
\ee
It is straightforward to check that the Bekenstein entropy in eq.~(\ref{Bek}) satisfies the first law of thermodynamics
\be
\label{firstlaw}
dM=TdS\;,
\ee
where the mass $M$ is determined by the opening angle
 \[
 M=2\pi M_3\l 1-\frac{\theta_o}{2\pi}\r=2\pi M_3(1-\epsilon^{1/2})\;.
 \]
Indeed, by definition 
$
f(R_0)=0
$,
so,  taking into account (\ref{energyconservation}), one obtains
\be
\label{de}
d\epsilon=-f'(R_0)dR_0\;.
\ee
Using (\ref{de}) one immediately finds that the first law of thermodynamics (\ref{firstlaw}) indeed holds.

The non-extremal solutions  take  an especially simple form in the limit  when the opening angle is so small 
that the radius of the compact dimension at the horizon is much larger than the mass scale of all massive particles.
In this limit
the Casimir energy is just
\[
\rho(R)=-\frac{\zeta(3)n_0}{(2\pi)^4R^3}\;,
\]
where $n_0$ is the total number of the massless degrees of freedom.
Plugging this Casimir potential  into (\ref{Ueff}) and (\ref{energyconservation}) and 
performing the rescalings $t\to\epsilon^{1/2}t$ and $R\to R\epsilon^{-1/2}$ one recognizes 
in the $(tR)$ part of the metric (\ref{3DU}) the radial part of the ($3+1$)-dimensional 
Schwarzschild metric with Schwarzschild radius 
\[
r_s=\frac{\zeta(3)n_0}{M_32\pi\theta_o^{3}}\;,
\]
 where the asymptotic opening angle  $\theta_o$ of the compact $\phi$-coordinate is related to the ``energy"
$\epsilon$ in the same way  as before, $\theta_o=2\pi\epsilon^{1/2}$.
 Unlike the charged extremal black hole these solutions do not have a smooth decoupling limit.
 Indeed, in the limit of large $M_3$ with fixed opening angle $\theta_0$ (so that the Bekenstein entropy remains finite), 
 the Hawking temperature $T_H=(4\pi r_s)^{-1}$ diverges and one cannot trust the semiclassical geometry.

Actually, such non-extremal quantum black holes were known before \cite{Emparan:1999wa,Emparan:2002px},
and were constructed in a way that provides a complementary viewpoint to understand their origin,
and simultaneously serves as a nice consistency check for our calculation.
Namely, the metric (\ref{3DU}) with  function $f(R)$ of the Schwarzschild form was found to describe
black holes localized on the Planck brane in the AdS$_4$ Randall--Sundrum setup. 
From the holographic CFT point of view these are black holes in three-dimensional gravity coupled to the large $N$ CFT. 
Now, at the classical level
there is no attractive force in three-dimensional gravity, and the only effect 
of the point mass on the geometry is to produce the conical deficit angle, so there can be no horizon.
This is no longer true at the quantum level; the one loop correction to the graviton propagator 
gives rise to an attractive potential \cite{Wessling:2001jb}
and as a result the existence of a horizon becomes possible. On the AdS side these quantum effects
are captured by the classical dynamics in the bulk, so that the induced metric on the Planck brane
indeed describes the quantum black hole geometry in the lower dimensional theory. Of course, the attractive 
one-loop potential is generated for a general matter sector as well, not just for the large $N$ CFT, and 
 ``Schwarzschild"  solutions can be  found in this way in the  purely three-dimensional setup as 
well (see, e.g. \cite{Soleng:1993yh}).

There is a little puzzle here---the one-loop correction to the graviton propagator 
leads to the attraction, independently of whether a particle circling around the loop is boson or fermion.
On the other hand, for the existence of the compactified vacuum and of the extremal interpolating solution 
 is crucial that fermions contribute to the Casimir energy with the opposite sign.
The resolution is related to the ${\mathbb Z}_2$ flux discussed above. In the absence of the flux, the fermions are 
antiperiodic and the one-loop potential is necessarily attractive.
Turning on the flux leads to periodic boundary conditions, making their contribution to the
one-loop potential repulsive.

Finally, there are also solutions with negative energy $\epsilon$. The  meaning of these geometries is 
not apparent
with the metric ansatz (\ref{3Dsimple}), as the only solutions of the mechanical problem that
reach the $R=\infty$ region  in this case are those with the imaginary
``time" $z$. However, presenting the metric in the form
(\ref{3DU}) makes it explicit that these are as meaningful solutions of the Einstein equations as those with 
positive energy $\epsilon$. Unlike the latter,  solutions  with negative energy do not asymptote to the conical
geometry in the asymptotically flat (large $R$) region. Instead, they describe  anisotropic cosmologies
with $R$ playing the role of time. In the large $R$ region they take the form
\[
ds^2=-\frac{dR^2}{|\epsilon|}+ |\epsilon|dt^2+R^2 d\phi^2\;.
\]
Locally this is just a Minkowski metric, with the $(R\phi)$ part of it being the expanding Milne universe.
Globally there is a difference from the Milne universe due to the compactness of the $\phi$-coordinate.

In  Fig.~\ref{fig:ads3D} we collected together the different options discussed above---large energies corresponding to the naked
conical singularities, critical energy $U(R_0)$ (extremal black hole), small positive energies 
(non-extremal black holes) and negative energies (cosmologies). We also presented  a schematic cartoon
of the geometry in each  case, and the corresponding Penrose diagrams.
In particular, we see that  the conformal diagram corresponding to solutions with negative energy
has the same form of the Penrose diagram for Schwarzschild black holes rotated by ninety degrees.
This diagram 
describes the anisotropic  bouncing cosmology, where the radius of the compact
dimension starts at infinity and bounces back. The scale factor $f(R)$ in front of the non-compact spatial
(in the asymptotically Minkowski region) coordinate $t$ bounces as well.
The big crunch/big bang singularity is partially resolved by the Casimir energy, in a sense that
observers can survive a transition from the contracting to the expanding stage without ever
hitting a time-like singularity at $R=0$. Note, that similarly to the inner horizon of 
the Reissner--Nordstrom black hole, the horizon replacing the big crunch singularity
suffers an instability with respect to the small perturbations of the initial data.

The quantum black holes that do not carry the ${\mathbb Z}_2$ flux are also straightforward to identify.
In this case fermions satisfy antiperiodic boundary conditions, so that their contribution to the Casimir
energy has the same sign as bosons.
The solutions of the corresponding mechanical problem  describe either bouncing cosmologies with
an unresolved singularity (imaginary time solutions with negative energies), 
or uncharged quantum black holes (positive energy solutions with a single turning point).

\begin{figure}[t]
\begin{center} \epsfig{file=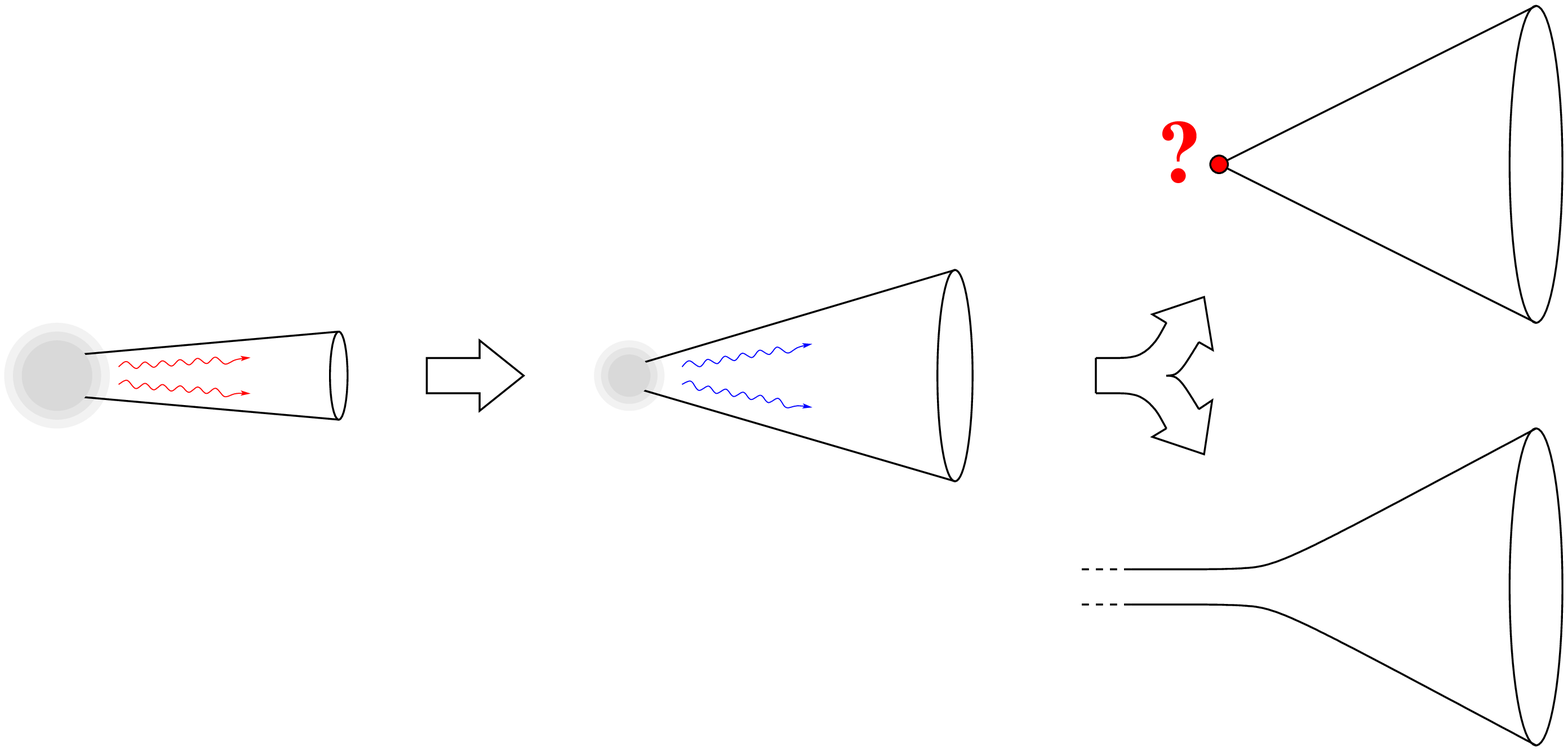,width=.8\textwidth} \end{center}
\caption{{\small \it As the quantum black hole evaporates its conical opening angle opens up. Depending  on the presence of 
 ${\mathbb Z}_2$ flux this process either stops at the critical 
 opening angle (\ref{eps}) or continues until the horizon shrinks  to the Planckian size.}}
\label{evolution}
\end{figure}

This discussion implies the following evolution history for the quantum black holes
 (see Fig.~\ref{evolution}),
after one  takes into account the Hawking evaporation.
One starts with a black hole of a  near critical Planckian mass, which is a very narrow cone with a singularity shielded by  
 the quantum horizon. As a result of the Hawking evaporation the horizon shrinks and the angle of the cone 
 opens up. Depending on whether the ${\mathbb Z}_2$ flux is present or not, this process either stops at the critical 
 opening angle (\ref{eps}) and the extremal black hole forms, 
 or continues until the cone opens completely and the horizon shrinks to the Planckian size.

\subsection{Interpolation from AdS$_3$ and dS$_3$ vacua}
There are no difficulties in extending the above discussion to the case when the three dimensional vacuum
is either AdS$_3$ or dS$_3$. All general results of the section~\ref{setting} still apply, the only difference
being that the effective potential does not vanish at $R=\infty$. 

For instance, in the
AdS$_3$ case the effective potential behaves as $U(R)\propto -R^2$.
 As before, the extremal interpolating geometry corresponds to the critical solution of the 
 mechanical problem with $\epsilon=U(R_0)$.  The large $R$ region of the metric (\ref{3DU})
asymptotes now to the boundary of the AdS$_3$.

 Just as in the flat case for larger values of 
$\epsilon$ one obtains AdS$_3$ geometries with a naked conical singularity, and for smaller values 
of $\epsilon$ non-extremal black holes. The only difference with the flat case is the absence
of the solutions that approach the asymptotically AdS region as cosmologies.

\begin{figure}
\begin{center} \epsfig{file=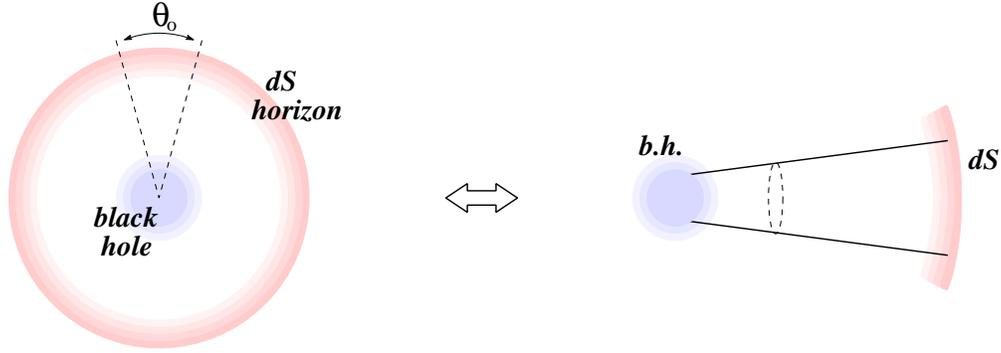,width=.8\textwidth} \end{center}
\caption{{\small \it Quantum black hole in dS$_3$ has a narrow opening angle
that cuts out most of the causal patch.}}
\label{twodesitters}
\end{figure}

In a sense, the  situation is the opposite in the case of the asymptotic dS$_3$ geometry.
Namely, in this case the potential of the mechanical problem is positive at large $R$, $U(R)\propto
R^2$, so that at any value of energy $\epsilon$ the large values of $R$ belong to the classically 
forbidden region of the auxiliary mechanical problem. In analogy to what we had at 
$\epsilon<0$ in section~\ref{nonextremal}, this implies that $R$ plays the role of time in this region,
so that the metric (\ref{3DU}) describes an inflationary three-dimensional Universe at the largest values of $R$.
A turning point of the mechanical solution at large $R$ corresponds to the horizon of the
static patch of dS$_3$. As before, at large values of $\epsilon$ the solution (\ref{3DU}) does not have 
any other horizons and develops a naked conical singularity at the origin  $R=0$ of the static patch. 
For extremal solutions with $\epsilon=U(R_0)$ this singularity is resolved into an infinite
AdS$_2\times$S$_1$ throat. At even smaller values of $\epsilon$ it is shielded by a non-extremal
horizon. Smallness of $\epsilon$ implies that these solutions contain only a tiny fraction of  the
de Sitter horizon, see Fig.~\ref{twodesitters}.

One peculiarity of the asymptotically dS$_3$ case, is the existence of a new extremum (minimum)
of the effective potential $U(R)$ at $R=R_1$. According to the discussion of section~\ref{setting} this
minimum corresponds to the dS$_2\times$S$_1$ vacuum. As a result, solutions (\ref{3DU})
with $\epsilon$ close to $U(R_1)$ develop a number of new interesting features. We will discuss these in section~\ref{todSsec}, 
where we describe interpolation to the dS$_2\times$S$_1$ vacua.

Here we would like to discuss another important question related to the interpolation from the higher
dimensional de Sitter space.  Namely, the finite entropy $S$ of the de Sitter horizon strongly suggests that 
the  Hilbert space describing possible quantum states of the de Sitter space is finite dimensional.
In turn, together with the thermal nature of the de Sitter vacuum, this implies that the de Sitter observer 
should go through all possible states before the Poincare recurrence time $\sim e^S$.

In the context of the metastable de Sitter vacuum, which can decay to the Minkowski or
AdS vacuum with the same number of spatial dimensions, this expectation is
supported by the remarkable fact that the semiclassical decay rate described either by the Coleman--de Luccia~\cite{CdL} or 
Hawking--Moss~\cite{HM} instantons is always faster than $\e^{-S}$, no matter how high
the barrier between the two vacua is. On the other hand, we have
not found solutions which have the interpretation of the expanding bubble of the lower
dimensional vacuum in the higher dimensional one. 
 How is that compatible with the argument sketched above, that the de Sitter space should
be able to populate all other states at the times scales shorter than the recurrence time?

The existence of the interpolating black hole solutions found here indicates that this process
is rather different from the conventional Coleman--de Luccia vacuum decay. Namely, instead of creating 
an expanding bubble of the new vacuum, de Sitter thermal fluctuations may lead to the collapse
of most part of the static patch into the quantum black hole described here. Afterwards
this black hole will  Hawking evaporate and approach an extremal interpolating solution.
We did not attempt to find an explicit instanton solution describing such a process. Such
an instanton may have rather peculiar properties, as it should change the value of the ${\mathbb Z}_2$ charge
within a causal patch. Note, that a creation of a pair of extremal black holes (such a configuration is neutral
with respect to ${\mathbb Z}_2$) is not possible within one causal patch, because each of the
black hole has a deficit angle close to $2\pi$. On the other hand, there is no conservation
law for the charge within a given causal patch and, consequently, no reasons to expect that such an instanton
does not exist. Note that unlike for the usual Coleman--de Luccia bubble this transition does not change the microscopic structure of the
vacuum, so that small enough observers (for instance, many of the {\it  Amoebozoa}) are able to  survive it.

\subsection{Interpolation in 4D}
Let us now discuss how the interpolating solutions look like in more realistic situations, namely let us describe
solutions interpolating from four- to three-dimensional vacua.  
For simplicity, we will mainly focus on the solutions interpolating from the four-dimensional Minkowski space to 
AdS$_3\times$S$_1$. As we discussed in section \ref{sec:SMlandscape} this case is relevant for the Standard Model neutrino
vacua, in the approximation when one neglects the effects related to the presence of the four-dimensional
cosmological constant. 
In this case we are looking for a 
cosmic string-like geometry, so that a natural generalization of the three-dimensional ansatz
(\ref{3Dinter}) is
\be
\label{4Dinter}
ds^2=A^2(z)\l- dt^2+dx^2\r +dz^2+R^2(z)d\phi^2\;,
\ee
where $x$ is the non-compact spatial coordinate along the string.
Note, that {\it a priori} there is no reason to assume Lorentz invariance in the $(tx)$ plane, as we did
in the ansatz (\ref{3Dinter}). As we will see, assuming this symmetry allows to obtain the extremal interpolating
geometry, while giving up this symmetry will lead to the related family of non-extremal black objects.
For simplicity let us proceed with the Lorentz invariant ansatz (\ref{3Dinter}).
The energy-momentum tensor still takes the form (\ref{TMN}), where now, of course, $\mu,\nu=t,x,\phi$.
The $(tt)$, $(zz)$ and $(\phi\phi)$ components of the Einstein equations then take the following form
\begin{eqnarray}
\label{4tt}
M_{4}^2\l R''+R'\frac{A'}{A}+R\frac{A''}{A}\r&=&-R\,\rho(R)\;,\\
\label{4zz}
M_{4}^2 \, \frac{A'}{A}\l 2R'+R\frac{A'}{A}\r&=&-R\,\rho(R)\;,\\
M_{4}^2\left [ \l\frac{A'}{A}\r^2+2\frac{A''}{A}\right]&=&-\left [ \rho(R)+R\,\d_R\rho(R)\right ]\;.
\label{4ff}
\end{eqnarray}
 To proceed it is convenient to solve for ${A'/A}$ from the $(zz)$-equation (\ref{4zz}),
\be
\label{ApA}
\frac{A'}{A}=-\frac{R'}{R}\pm\sqrt{\l \frac{R'}{R}\r^2-\frac{\rho}{M_{4}^2}}\;.
\ee 
 To understand the meaning of the sign ambiguity in (\ref{ApA}), note that the asymptotically flat boundary conditions
 at $z=-\infty$ are
 \[
 \left.\frac{R'}{R}\right|_{z=-\infty}=\frac{1}{z}<0\;,\;\;  \left.A'\right|_{z=-\infty}=0\;.
 \]
 These correspond to the ``$-$" sign in (\ref{ApA}) (recall, that we are assuming zero cosmological constant, so that
 $\rho(R)\to 0$ at large $R$). On the other hand, asymptotically flat boundary conditions
 at $z=+\infty$ require $R'$ to be positive and correspond to the ``$+$" sign in (\ref{ApA}).
 The existence of two branches in (\ref{ApA}) indicates that, just as in the three dimensional case, it is impossible to find a smooth solution
 of the form (\ref{4Dinter})  connecting
 two asymptotically non-compact flat regions at $z=\pm\infty$ (such a solution would be a Lorentzian wormhole).  In what 
 follows we choose the sign ``$-$" in (\ref{ApA}) so that the asymptotically flat region is at $z=-\infty$ (this convention
 is opposite to the one used before, however it is more convenient for the purposes of the present 
 discussion).
 Then one can take the combination of the $(tt)$ and $(\phi\phi)$ equations
 (\ref{4tt}), (\ref{4ff}) that does not contain $A''$, and
 plug (\ref{ApA}) there. As a result one arrives at the following equation for the
 radius of the compact dimension alone,
 \be
 \label{Req}
 R''+\gamma R'=-\d_R U\;,
 \ee
 where the effective potential $U$ is determined by
 \be
 \label{effpot}
\frac{d U(R)}{dR}=\frac{1}{M_{4}^2} R(\rho-R\,\d_R\rho)
 \ee
 and the friction parameter $\gamma$ is
 \be
 \label{fric}
 \gamma=-2\l\frac{R'}{R}+\sqrt{\l\frac{R'}{R}\r^2-\rho}\r\;.
 \ee
 The shape of the effective potential $U$ due to the Casimir energy in a theory with the light
 spectrum of the  Standard Model (and with zero cosmological
 constant) is the same as the one in Fig.~\ref{fig:ads3D}. As before, the maximum at $R=R_0$ corresponds to the compactified
 AdS$_3\times$S$_1$ vacuum and we are interested in the solution that starts at $R=\infty$ and makes it to the top
 of the potential in a infinite time. 
 
 The difference with the three-dimensional case is the presence of the
 friction term in (\ref{Req}).  It is straightforward to check that $\rho<0$ in the whole region
 to the right of the maximum, $R>R_0$, so that the friction parameter $\gamma$ is negative there and gives rise to
an  antifriction. The presence of this antifriction does not prevent us from running the argument proving
the existence of the extremal solution. Just like in the three-dimensional case in the limit of a very small opening
angle, $R'(-\infty)\to 0$, the solution to the mechanical problem (\ref{Req}) undershoots the maximum, while
for large opening angles it overshoots, so there is a critical value such that $R(z)$ monotonically 
drops down and stops at  $R_0$ in an infinite time.
From (\ref{ApA}) one sees that the warp factor $A(z)$ also monotonically drops down for this solution without ever 
changing its sign (recall, that 
we chose the ``$-$" sign in (\ref{ApA})) and at large $z$ approaches zero as
\[
A(z)\sim \e^{-z\sqrt{-\rho}}\;,
\] 
so the extremal solution indeed interpolates to the AdS$_3\times$S$_1$ vacuum.
On dimensional grounds it is clear that the asymptotic opening angle for the solution interpolating to the
neutrino vacuum of the  Standard Model  is
\[
\theta_o=2\pi |R'(-\infty)|\sim \frac{m_\nu}{M_{4}}\;.
\]
Similarly to  the three-dimensional case, solutions with larger opening angles overshoot and develop
a conical singularity. On the other hand, the behavior of the solutions with smaller opening angles
is different from the lower dimensional case. Namely, as one can see from (\ref{ApA}), 
the turning point $R'=0$ does not correspond to a horizon any longer, so the undershooting solutions
do not describe the non-extremal black strings. 
What happens instead is that, due to the presence of the antifriction term in (\ref{Req}), 
the radius of the compact dimension diverges at a finite distance after the turning point, so that the solution develops a naked singularity.
As we said before, in order to obtain the black non-extremal solutions one has to give up with Lorentz invariance in the
ansatz (\ref{4Dinter}).

The extension of these results to the AdS$_4$ case is straightforward. 
The only subtlety is that the asymptotically AdS$_4$ boundary condition at $z=-\infty$ implies that
$A\propto R\propto \exp(|z|/l_3)$, so that $R'$ is infinite. This makes it inconvenient to use $R$ itself as a variable in the
auxiliary mechanical problem. Changing variable to  $k=\log R$ in (\ref{Req}), one can literally repeat
the above argument to prove that the interpolating solution of the form (\ref{4Dinter}) exists in this case as well.
This solution can be interpreted as a holographic RG flow of a CFT$_3$ broken by compactifying one of the spatial dimensions on a circle to a CFT$_2$ in the IR. 

However, unlike in the lower-dimensional case, the ansatz (\ref{4Dinter}) is not the appropriate one to describe an interpolation 
from dS$_4$. A fast  way to see this, is to note that translational invariance in $x$ 
is incompatible with dS$_4$ symmetries. To see this explicitly it is enough to solve eqs.~(\ref{4tt}), (\ref{4zz}) and (\ref{4ff}) for a pure 
cosmological constant, $\rho(R)=const>0$. It is straightforward to check that the resulting vacuum solutions are never maximally symmetric,
{\it i.e.} dS$_4$ metric cannot be presented in the form (\ref{4Dinter}).  Instead,
the cosmic string geometry in the static dS$_4$ coordinates takes the form
\[
ds^2=-(1-r^2)dt^2+\frac{dr^2}{1-r^2}+r^2\l d\theta^2+\epsilon\sin^2\theta d\phi^2\r\;,
\]
where $\epsilon$ determines the deficit angle. It is likely that the problem of finding an interpolation between this
geometry and the AdS$_3\times$S$_1$ vacuum cannot be reduced to ordinary differential equations
and requires the analysis of a two-dimensional system of
partial differential equations with non-trivial dependence on both $r$ and $\theta$. Having seen how it works in three dimensions, 
in principle there should be no obstruction for the existence of the quantum black strings in dS$_4$.

\subsection{Interpolation to dS vacua}
\label{todSsec}
So far we focused on interpolations to low dimensional vacua with a negative cosmological constant. 
This situation is similar to the ordinary Reissner--Nordstrom black holes and is relevant for the 
neutrino vacua of the Standard Model (assuming neutrinos are Majorana). However it is  interesting
to consider also what happens when the lower dimensional vacuum has a positive cosmological
constant. 

Natural interpolating solutions in this case are Coleman--de Luccia bubbles describing
decompactifications of the lower dimensional vacua as discussed  in \cite{Giddings:2004vr}. Of course,
our four-dimensional Universe could not have originated from one of the Standard Model 
three-dimensional vacua in this way, as the reheating temperature would be too low. 
However, it would be interesting to study the observational cosmological consequences of the scenario where our 
Universe was created as a result of the decompactification of a lower dimensional metastable vacuum.
We will not address this issue here. 

Instead, given that in the three-dimensional setup we have an explicit solution (\ref{3DU})
that applies to the low dimensional de Sitter vacuum as well, let us discuss its properties in this case.
Note that compactifications to two dimensions are somewhat subtle because the radion field
is not dynamical. Nevertheless, as discussed in Appendix~\ref{2Dvacua}, there is a sense in which the de Sitter vacuum always 
corresponds to the maximum of the radion potential in this case.
Due to the absence of the dynamical radion this vacuum is classically stable under
local perturbations (actually, even in four dimensions a de Sitter maximum can be effectively stable if the
radion is light enough, so that the Universe is eternally inflating on ``the top of the hill").
As a result, instead of the Coleman--de Luccia type of bubbles one may expect the interpolating solution to describe
just a classical rolling from the top of the potential in this case.
\begin{figure}
\psfrag{Rinf}{\hspace{-2pt} \footnotesize \textcolor{red}{$R$=$\infty$}}
\psfrag{R0}{\hspace{-7pt} \footnotesize \textcolor{red}{$R$=$0$}}
\psfrag{BH}{\footnotesize BH}\psfrag{dS}{\footnotesize dS$_2\times$S$_1$}
\psfrag{bounce1}{\footnotesize \it Bouncing}
\psfrag{bounce2}{\footnotesize \it Cosmology}
\psfrag{R}{\small $R$} 
\psfrag{UR}{\hspace{-14pt} \small $U(R)$}
\psfrag{I1}{\footnotesize \bf \textcolor{blue}{I}}
\psfrag{I2}{\footnotesize \bf \textcolor{blue}{II}}
\psfrag{I3}{\footnotesize \bf \textcolor{blue}{III}}
\psfrag{I1p}{\footnotesize \bf \textcolor{blue}{I$'$}}
\psfrag{I2p}{\footnotesize \bf \textcolor{blue}{II$'$}}
\psfrag{I3p}{\footnotesize \bf \textcolor{blue}{III$'$}}
\begin{center} 
\begin{tabular}{ccc}
\epsfig{file=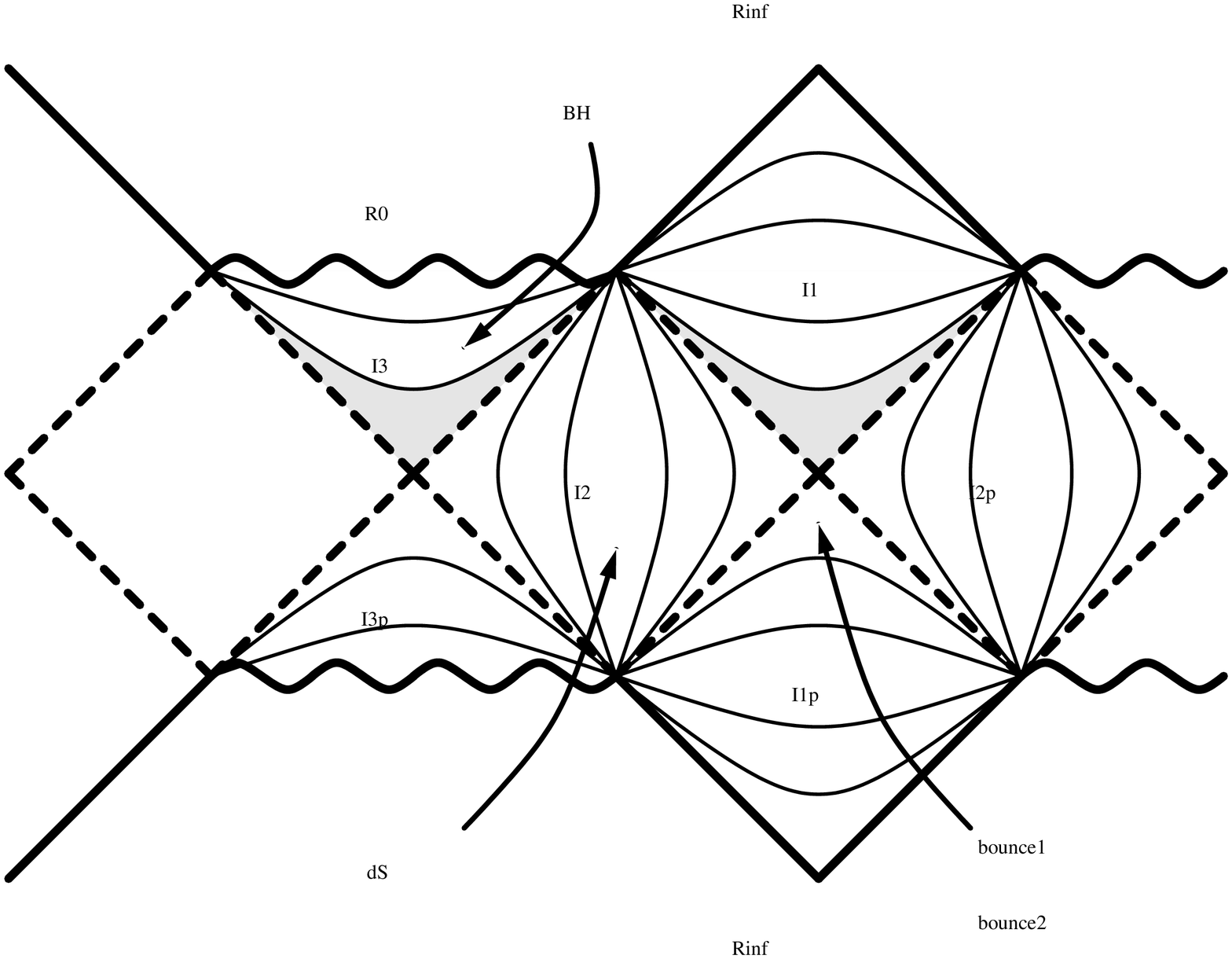,height=5cm} & $\mbox{}$\hspace{1cm}$\mbox{}$&\epsfig{file=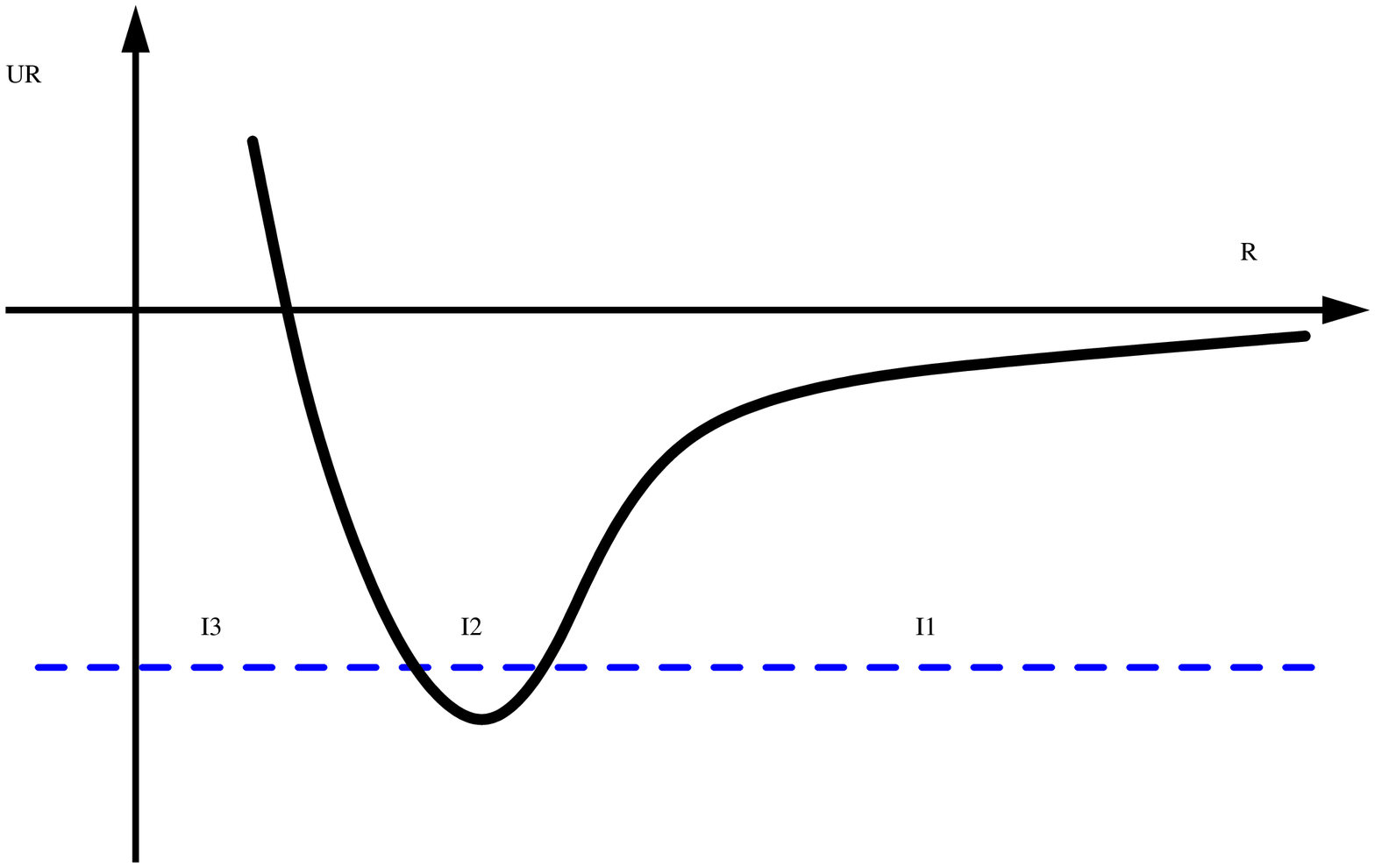,height=5cm} \\
a) && b)
\end{tabular}
\end{center}
\caption{{\small \it {\rm a)} Conformal diagram of the near extremal solution interpolating to the dS$_2\times$S$_1$ vacuum
and {\rm b)} the corresponding mechanical potential $U(R)$.}}
\label{fig:PentodS}
\end{figure}

The shape of the effective potential $U$ corresponding to the dS$_2\times$S$_1$ vacuum is shown in Fig.~\ref{fig:PentodS}b. 
It is straightforward to analyze the structure of the interpolating solutions (\ref{3DU}) at different values of $\epsilon$.
In all cases the corresponding Penrose diagrams are ninety degrees rotations of those shown in Fig.~\ref{fig:ads3D}.
Let us discuss here the solution exhibiting the richest pattern of features, namely the  near extremal one, with $\epsilon$ being
slightly larger than the value of the potential $U$ at the minimum. The corresponding Penrose
diagram is shown in Fig.~\ref{fig:PentodS}a. 

Part {\bf II} (as well as its  horizontally translated cousins {\bf II$'$} \dots) of this diagram 
corresponds to the classically allowed region of the mechanical problem. According to the 
discussion at the end of section~\ref{setting}, in the near extremal limit, the geometry of this region  is that of the causal diamond of the dS$_2\times$S$_1$ vacuum.
As usual, 
after continuation through the horizon the $R$-variable becomes time-like, while the $t$-variable
is space-like. So parts {\bf I} and {\bf III} of the Penrose diagram cover  regions with anisotropic 
cosmological expansion. 

The geometry of the region {\bf III} has a structure somewhat similar to that of the interior of the Schwarzschild black hole.
Namely, the compact coordinate shrinks in this region down to zero size
at the $R=0$ singularity. So, as a result of the quantum effects the conical singularity is replaced by a
big crunch singularity for the compact dimension. However, the function $f(R)$ grows indefinitely in this region, 
implying that the non-compact space-like coordinate $t$ experiences  superaccelerated cosmological expansion and
eventually hits the big rip singularity at $R=0$. 

One interesting difference with the black hole interior is due to the part of the region {\bf III} adjacent to the region
{\bf II} (grey shaded region in Fig.~\ref{fig:PentodS}a), where the effective potential is still approximately quadratic, 
\[
U\approx U(R_0)+\ell_2^2 (R-R_0)^2+\dots\;.
\]
Plugging this expression into $f(R)$, one finds that this part of the region {\bf III} is  an exponentially inflating
two-dimensional Universe in the FRW coordinates.
The size of the compact dimension slowly rolls down  here, so in a sense the radion plays the role of the
inflaton. This interpretation is somewhat subtle though, because, at least at the classical level, 
there is no dynamical radion in the compactifications  from three to two dimensions. 

Finally, it is  the existence of the region {\bf I} which signals that we are dealing with an interpolating geometry.
Indeed, in this region, the coordinate $R$ is also time-like and, as it grows to infinity, the function $f(R)$
approaches a constant value $f(+\infty)\approx -|U(R_0)|$, so that the metric is flat with the $(R\phi)$ part 
being the expanding Milne universe.
Globally there is a difference from the Milne universe due to the compactness of the $\phi$-coordinate.

Just like in the region {\bf III}, a shaded part of the region {\bf I} 
describes an exponentially inflating two-dimensional Universe.
Finally, regions {\bf I$'$, III$'$} describe the same cosmological solutions as   {\bf I, III} with the reversed direction of time.
In the vicinity of the boundary with the region {\bf II}  regions {\bf I$'$, III$'$} describe collapsing cosmologies and their horizons are very much
similar to the inner Cauchy horizon of the Reissner--Nordstrom black hole. As usual, such a horizon is unstable with respect to small
perturbations of the initial data, so regions {\bf I$'$, III$'$} are not to be there in a physically realizable situation. 

Consequently, as expected, the physical meaning of the
dS$_2\times$S$_1$ interpolating solution is to describe the inflation ``on the top of the hill" in the lower dimensional vacuum, which
ends up either in the singularity, where the compact dimension collapses, or exits into the asymptotically flat decompactified 
space-time. It will be interesting to calculate the spectrum of cosmological perturbation for this inflation.
As we already mentioned, a peculiar feature of this case is that there are no propagating perturbations of the
inflaton (radion) at the classical level. However, at the one-loop level we expect the inflaton to become dynamical.


%
%
%
%
%
%
%
%
%
%
\section{Conclusions}
We have seen that the Standard Model has a near-moduli space of 
lower-dimensional vacua with moduli stabilized by a combination of a tiny 
tree-level contribution from the cosmological constant and one-loop 
corrections. For the minimal theory of neutrino masses, there are AdS$_3\times$S$_1$ vacua, implying the existence of a dual CFT$_2$ describing 
the Standard Model coupled to Quantum Gravity. We also showed quite 
generally that it is possible to interpolate to lower-dimensional 
AdS vacua as near-horizon regions of new kinds of quantum extremal black 
objects---black strings in going from $4 \to 3$ dimensions, black holes 
from $3 \to 2$ dimensions. The extremal 3D black holes are 
particularly interesting---they are metastable objects with an entropy that 
is independent of $\hslash$ or $G_N$, so a non-gravitational microscopic 
accounting of their entropy might be possible in a decoupling limit where 
$G_N,\hslash \to 0$ and the geometry degenerates to a cone with a 
fixed, small opening angle.


There are a number of obvious issues that require further elaboration. 
We did not study the radion effective potential for radii smaller than the QCD scale, so we don't know if there are additional 
vacua there. Nor have we analyzed the SM potential 
in the case of even lower dimensional compactifications.
It would be interesting to explicitly find the gravitational solutions that interpolate between dS$_4$ and 
3D vacua---symmetry considerations suggest that the problem is different from its lower-dimensional analogue. 
We also did not attempt to find interpolating solutions from 4D to 2D vacua. 
If 3D de Sitter vacua can exist, it is natural to ask if our universe could have originated from tunneling out of eternal inflation in 3D. Of course  
we need to have a phase of slow-roll inflation {\em after} the nucleation of our 4D bubble takes place, so the tunneling 
should happen with the inflaton stuck at the top of its potential. It would be interesting to investigate 
possible cosmological signatures of such a scenario. Finally, we have not explicitly constructed instantons for transitions between deSitter space and the extremal quantum black objects.


A crucial ingredient for both the existence of these non-SUSY vacua and 
the quantum horizons allowing interpolation is the violation of the 
null-energy condition and negative gravitational energy associated with 
the Casimir effect. It is interesting that objects with negative 
gravitational energy play a crucial role in {\it all} modern mechanisms 
for stabilizing moduli to flat or dS spaces such as KKLT~\cite{KKLT}; for instance negative tension orientifold planes are 
present in these constructions. Just as 
the new Standard Model vacua we have found are associated with quantum 
black objects, it is natural to conjecture that at least the AdS vacua 
in the string landscape can be realized as near-horizon geometries of new 
black brane solutions asymptoting to 10 or 11 dimensions, or more 
generally some point on the maximally supersymmetric moduli space. The 
orientifolds must play a crucial role in allowing the existence of these 
solutions. The landscape of lower-dimensional vacua should thus be 
associated with a zoo of exotic black hole solutions, allowing us to look at the vacua from the ``outside". 
It would be interesting to 
try and find these black brane solutions explicitly for the classical IIA 
vacua of \cite{IIA}. As a simpler warm-up with the same essential 
features---negative tension and fluxes--- consider stabilizing a 1D interval 
(or S$_1/{\mathbb Z}_2$ orbifold), by having a negative tension $T$ on one end of the 
interval and an axion with decay constant $f$ and fixed periodicity around 
the circle. Such a situation could well exist for our vacuum; if there is 
low-energy SUSY, we could have $T \sim -m_{SUSY}^4$ and the QCD axion 
suffices. The radion effective potential is $V_{eff}(R) \sim R^{-3} 
(\Lambda R -|T| + f^2/R)$; $\Lambda$ is negligible here and there is a 
non-trivial AdS minimum. The interpolating geometry in this case should 
look like a narrow strip, bounded by the negative tension brane on one end 
and the other end of the interval on the other, again with a small opening 
angle.

The necessity of negative energy objects in realistic models of modulus 
stabilization has sometimes been thought of as a technicality---but we have 
seen that they are associated with new sorts of horizons and thus 
surprising causal structures in the higher-dimensional geometries the 
lower-dimensional vacua are embedded in. It is worth exploring this issue 
further. For instance, we often imagine tunneling out of stabilized dS 
vacua to 10/11 dimensional supersymmetric space-times; but this is not 
correct. The asymptotic spaces must not only carry a remnant of e.g. the 
fluxes labeling the vacua, but they also have e.g. orientifold planes with 
negative gravitational energy. How do these affect the geometry?

\section*{Acknowledgments}
We thank Tom Banks, Raphael Bousso, Michael Dine, Gia Dvali, Juan Maldacena, Nathan Seiberg, Marco Serone, Steve Shenker, Andy 
Strominger, Raman Sundrum, Cumrun Vafa and Edward Witten for stimulating discussions.
Our work is supported by the DOE under contract DE-FG02-91ER40654. 

\appendix

\section{Casimir Energy}
\label{app:Casimir}
In this appendix we review the derivation of the 1-loop Casimir contribution to
the energy-momentum tensor for a generic massive field, in $d$-dimensions with 
one dimension compactified on a circle. Let us call $x^d=y\in [0,2\pi R)$ 
the compact dimension. Given a free scalar field with Lagrangian
\beq
{\cal L}=-\frac12 \l \de \Phi \r^2-\frac12 m^2 \Phi^2\,, \nn
\eeq
at 1-loop the expectation value of the energy-momentum tensor reads
\bea 
\big \la T_{\mu\nu}\big \ra &=&\left \la {\cal L}\, g_{\mu\nu}-2 \frac{\delta {\cal L}}{\delta g^{\mu\nu}} \right \ra \nn  \\
&=& \lim_{x'\to x}\left [ \frac12 \l \de_\mu \de'_\nu +\de_\nu \de'_\mu \r -\frac12 g_{\mu\nu} \l \de^\rho \de'_\rho+m^2\r\right ]
G(x-x')\,, \label{eq:Tmncas}
\eea
where $G(x-x')=\la \Phi(x) \Phi(x')\ra$ is the free propagator.
When one dimension is compact the Casimir contribution can easily be obtained
just by summing the infinite volume Green function over all the images, namely
\beq
G(x-x')= {\sum}'_n G_\infty(x-x'+2\pi R\, n\, \hat y)\,, \nn
\eeq
in the sum $n$ runs over all integers but $0$, which corresponds to the infinite volume
$R$-independent contribution that must be reabsorbed into the cosmological constant.
Notice that, having subtracted the $n=0$ contribution, also the second term in eq.~(\ref{eq:Tmncas}) vanishes.
So we finally have
\bea
\big \la T_{\mu\nu}\big \ra &=&\frac12 \lim_{x'\to x} \l \de_\mu \de'_\nu +\de_\nu \de'_\mu \r {\sum}'_n G_\infty(x-x'+2\pi R\, n\, \hat y) \nn \\
&=&-\, {\sum}'_n \left. \de_\mu \de_\nu G_\infty(y_n )\right |_{y_n=2\pi R n \hat y} \nn \\
&=&- \left[ \rho(R)\, \eta_{\mu\nu} +R \,\rho'(R)\,\delta_\mu^y\,\delta_\nu^y \right]\,, \label{eq:Tmunu}
\eea
where
\beq 
\rho(R)=2\,{\sum}'_n \left. \frac{\de  G_\infty(y_n^2)}{\de y_n^2}\right |_{y_n=2\pi R\, n\, \hat y}\,, \nn
\eeq
is the Casimir energy density. 
In the case of charged fields we can also have non-periodic boundary conditions
\beq
\Phi(x,y+2\pi R)=e^{i \theta}\Phi(x,y)\,, \nn
\eeq
and the Green functions in the sum get an extra Wilson line contribution
\beq
{\sum_n}' e^{i n \theta} G_\infty(x-x'+2\pi R n \hat y)\,. \nn
\eeq
So the final expression for the Casimir energy density in the general case reads
\beq
\label{eq:rhoCas}
\rho(R)=2\,{\sum}'_n \left. e^{in\theta} \frac{\de  G_\infty(y_n^2)}{\de y_n^2}\right |_{y_n=2\pi R\, n\, \hat y}\,.
\eeq
This formula applies also for fermion, vector and graviton fields, with an extra minus in the case of fermions.
By plugging in the explicit formula for the Green function one can easily read the result.
For example, in the case of $d=4$ for a massless field with periodic boundary conditions, the Green function reads
\beq
G_\infty(y_n^2)=\frac{1}{4\pi^2 y_n^2}\,, \nn
\eeq
so that eq.~(\ref{eq:rhoCas}) gives
\beq
\rho(R)=-\frac{4}{(2\pi)^6 R^4}\sum^\infty_{n=1} \frac{1}{n^4}=-\frac{\pi^2}{90}\frac{1}{(2\pi R)^4}\,. \nn
\eeq
The contribution in the effective potential in the dimensionally reduced $3D$ theory
reads
\beq
V_C= 2 \pi R\, \rho(R)=-\frac{1}{720 \pi R^3}\,, \nn
\eeq
while the contribution in the Weyl-rescaled metric of eq.~(\ref{eq:4to3metric}) is just
\beq
-\frac{r^3}{720 \pi R^6}\,. \nn
\eeq

From the form of the energy-momentum tensor (\ref{eq:Tmunu}) we can easily derive 
the condition for $\rho(R)$ not to violate the Null Energy Condition
\beq
T_{\mu\nu} n^\mu n^\nu \geq 0 \,, \qquad \forall n^\mu: n^2=0\,, \nn
\eeq
and reads
\beq
T_{\mu\nu} n^\mu n^\nu = - 2 (n^y)^2 R \,\rho'(R) \geq 0\,, \nn 
\eeq
\beq
\rho'(R)\leq 0\,, \nn
\eeq
which is satisfied by fermions but violated by bosons. 

Let us now derive the explicit formula for $\rho(R)$ in the most general case.
Since we are interested to the value of the Green function outside the light-cone
we can work directly in Euclidean space, the Green function then reads
\beq
G_\infty(x)=\int \frac{d^d k}{(2\pi)^d} \frac{e^{i k x}}{k^2+m^2}
=\frac{m^{d-2}}{(2\pi)^{d/2}} \frac{K_{d/2-1}(m\, x)}{(m\, x)^{d/2-1}}\,, \label{eq:Gfgen}
\eeq
where $K_\nu(z)$ is the Bessel function
\beq
K_\nu(z)= \frac12 \int_0^\infty d\beta\, \beta^{\nu-1} \ e^{-\frac{z}{2}\l\beta+\frac{1}{\beta}\r}\,. \nn
\eeq
Now, by using the fact that
\beq
\de_z \l \frac{K_\nu(z)}{z^\nu}\r=-\frac{K_{\nu+1}(z)}{z^\nu}\,,\nn
\eeq
inserting the result for the Green function (\ref{eq:Gfgen}) into eq.~(\ref{eq:rhoCas}) we get
\beq
\rho(R)=-\sum_{n=1}^{\infty} \frac{2 m^d}{(2\pi)^{d/2}} \frac{K_{d/2}\l2\pi R\, m\, n\r}{(2\pi R\, m\, n)^{d/2}}\cos(n\,\theta)\,. \nn
\eeq
The massless limit can easily be taken by noticing that for $z\to0$,
\beq
z^\nu K_\nu (z)=2^{\nu-1} \Gamma(\nu) \left[ 1 -\frac{z^2}{4(\nu-1)}+O(z^4)\right]\,, \nn
\eeq
and reads
\beq
\rho(R) = -\frac{2}{(2\pi R)^d \Omega_{d-1}}\, \Re\,\left[ 
{\rm Li}_{d} (e^{i \theta}) - \frac{2 \pi^2\, {\rm Li}_{d-2} (e^{i \theta})}{d-2}  (m R)^2 + O(m R)^4 \right]\,, \nn
\eeq
where 
\beq
{\rm Li}_n (z)\equiv \sum_{k=1}^{\infty} \frac{z^k}{k^n}\,, \qquad 
\Omega_{d-1} \equiv \frac{2\pi^{d/2}}{\Gamma(\frac{d}{2})}\,, \nn
\eeq
${\rm Li}_n (1)=\zeta(n)$, ${\rm Li}_n (-1)=(2^{1-d}-1)\zeta(n)$ and $\zeta(n)$ is the Riemann zeta-function.
Notice also that the first corrections to the massless limit is negative and proportional
to $(m\,R)^2$.

Analogously for $m\to \infty$, using 
\beq
z^\nu K_\nu (z)\xrightarrow{ z\to \infty } \sqrt\frac{\pi}{2}\, z^{\nu-\frac12}\, e^{-z} \,, \nn
\eeq
we get
\beq \label{eq:rhoasympt}
\rho(R)\xrightarrow{m\to \infty} -\frac{(mR)^{\frac{d-1}{2}}}{(2\pi R)^d}\,e^{-2\pi R m}\,\cos(\theta) \,,
\eeq
which shows the exponential suppression for $m R>1$.

\section{More Vacua}
\subsection{Other 3D SM vacua}
\label{app:more3Dvacua}
In section~\ref{sec:SMlandscape} we showed how Casimir contributions to the effective
potential of the radion may determine a non trivial vacuum, actually a continuum, at
the micron scale. One can now ask what happens at shorter distances.
For smaller sizes of the radius the neutrinos are effectively massless
and since the number of fermionic degrees of freedom is larger than the
number of bosonic ones, with periodic boundary conditions the scalar potential
grows, independently of the value for the Wilson loop.
Nothing new happens until the size of the radius approaches the 
Compton wavelength of the electron. At this point also the electron d.o.f.
start to be important. Moreover, since the electron is charged,
also the Wilson loop will start receiving important contributions:
for $\theta=0$ the contribution to the effective potential is positive
and it continues to grow; For $\theta=\pi$, on the other hand, 
the contributions from the fermions is negative, the potential starts
decreasing, developing a saddle point at $R\sim 1/m_e$ and $\theta=\pi$. 
It seems that the structure of the SM potential is getting more and more interesting.

Because in three dimensions the electromagnetic coupling is relevant, one could
worry that at large distances the theory becomes strongly coupled
and the calculation breaks down, however, it is easy to check that,
as long as the 4D coupling is perturbative, this happens only at distances
parametrically larger than the radius, and the calculation is always
within the regime where it can be trusted. 

For smaller radii more and more states come in, changing at each stage
the behavior of the potential. 
If we define the single bosonic contribution to the Casimir energy as
\beq
V_C^{(1)}[R,m,\theta]\equiv - \frac{r^3\, m^4 }{\pi R^2} \sum_{n=1}^\infty 
\frac{\cos(n \, \theta)}{(2 \pi R\, m\, n)^2} K_{2}\l 2 \pi R\, m\, n \r\,, \nn
\eeq
the full effective potential will then read
\beq \label{eq:Vpot}
V=\frac{2\pi r^3 \Lambda_4}{R^2}+\sum_{a} (-1)^{F_a} n_a\, V_C^{(1)}\left [ R,\,m_a,\,2 \pi \l q_a A_\phi+\frac{1-z_a}{2}\r \right ]\,,
\eeq
where: the sum goes over the whole SM spectrum from massless states to the QCD pseudo-Goldstone bosons 
(after which the theory becomes non-perturbative), $F_a=0,1$ if the $a$-th state is bosonic or fermionic respectively,
$n_a$ counts the d.o.f. of the $a$-th state (1 for scalars, 2 for massless vectors, 4 for Dirac fermions\dots), 
$m_a$ is the mass, $q_a$ is the absolute value of the
electric charge normalized to that of the electron $e$, $A_\phi$ is the Wilson loop modulus and $z_a=0,1$
for periodic or antiperiodic boundary conditions.

Because of the asymptotic behavior of $V_C$ (eq.~(\ref{eq:rhoasympt})),
as long as $R$ is away from threshold regions ($\sim 1/m_a$) the total contribution to $V$ is just the sum
of the massless contributions from states that are lighter than $1/R$. All these contributions are the same
up to a constant factor that depend on the number of d.o.f., the periodicity of the field (also due to a non-trivial
Wilson loop) and on the fermionic number of the state ($F_a$). Just looking at these factors one can check
the overall sign of the contribution for each $R$, which determines the derivative of $V$ with respect to $R$,
thus the presence of stationary points. In table~\ref{tab:caspot} we reported such counting for periodic
boundary conditions, which shows that besides the neutrino vacuum and a saddle point at the electron scale
no other stationary points show up until $R\approx \Lambda_{QCD}^{-1}$.

\begin{table}
\begin{center}
\begin{tabular}{|c | c || c  | c | c | c | c | c | c |}
\hline
& $\theta$ & $g,\gamma$ & $\nu$ & $e^-$ & $\mu^-$ & $\pi$ & $K$ & $\eta_8$ \\
\hline \hline M
& $0$   & $-4$ & $2$ & $6$ & $10$ & $7$ & $3$& $2$ \\ M
& $\pi$ & $-4$ & $2$ & $-3/2$ & $-5$ & $-17/4$ & $-9/2$ & $-11/2$ \\
\hline \hline  D
& $0$   & $-4$ & $8$ & $12$ & $16$ & $13$ & $9$& $8$ \\ D
& $\pi$ & $-4$ & $8$ & $9/2$ & $1$ & $7/4$ & $3/2$ & $1/2$ \\ 
\hline
\end{tabular}
\end{center}
\caption{{\small \it Total number of d.o.f. after each threshold weighted with the factors $\mp1$ for bosons or fermions
and with $1$ or $-7/8$ for charged fields if the Wilson loop value is $\theta=0$ or $\pi$ respectively.
The two cases refer to Majorana (M) and Dirac (D) neutrino. A change in sign signals a stationary point.}}
\label{tab:caspot}
\end{table}
At this point the perturbation theory breaks down and we cannot trust the formula
for the potential (\ref{eq:Vpot}) anymore.
In order to study the radion potential around the QCD scale  one would need a non-perturbative analysis,
using, for instance, lattice QCD simulations. So at the moment we cannot say whether other
SM vacua are present in this region for the radion.
However, at smaller distances, the strong interaction  becomes weak and we can restart
using perturbative formulae for our study. This times counting the elementary d.o.f.: gluons, quarks\dots

At this point, however, the structure of the effective potential gets much more involved.
First of all, quarks bring fractional charges that, at fixed radius, 
produce more than one local minima for the Wilson loop.
Second, also gluons can develop non-trivial Wilson loops. There are
actually two more moduli ($G_\phi^{(1,2)}$) to be considered, associated
to the generators of the Cartan subalgebra of $SU(3)$. Both
quarks and gluons generate, at the quantum level, non-trivial contributions to the
scalar potential for these two fields.
If one, or both of them, develop a non-vanishing expectation value
than the $SU(3)$ color group breaks spontaneously into $SU(2)\times U(1)$
or $U(1)\times U(1)$. The effective potential now read
\beq
V=\frac{2\pi r^3 \Lambda_4}{R^2}+\sum_{a} (-1)^{F_a} n_a V_a\,, \nn
\eeq
where for gluons
\bea
V_a&=&2\,V_C^{(1)}\left [ R,\,0,\, 0 \right]+2\,V_C^{(1)}\left [ R,\,0,\,2 \pi \l G^{(1)}_\phi-G^{(2)}_\phi \r \right ] \nn \\
&&\mbox{}+2\,V_C^{(1)}\left [ R,\,m_a,\,2 \pi \l 2\,G^{(1)}_\phi+G^{(2)}_\phi \r \right ]
+2\,V_C^{(1)}\left [ R,\,m_a,\,2 \pi \l G^{(1)}_\phi+2\,G^{(2)}_\phi \r \right ]\,, \nn
\eea
while for the other fields:
\bea
V_a&=&V_C^{(1)}\left [ R,m_a,2 \pi \l q_a A_\phi+g\,G^{(1)}_\phi+\frac{1-z_a}{2}\r \right ] \nn\\
	&&\mbox{}+g\,V_C^{(1)}\left [ R,m_a,2 \pi \l q_a A_\phi+g\,G^{(2)}_\phi+\frac{1-z_a}{2}\r \right ]\nn\\
	&&\mbox{}+g\,V_C^{(1)}\left [ R,m_a,2 \pi \l q_a A_\phi-g\,G^{(1)}_\phi-g\,G^{(2)}_\phi+\frac{1-z_a}{2}\r \right ]\,, \nn
\eea
where $g=1$ (or $0$) if the field is (or is not) a quark.
The potential became a highly non-trivial function of the radion and the three Wilson loops ($A_\phi$, $G_\phi^{(1,2)}$)
and the search for stationary points becomes much more involved.

Above the weak scale one would need to know also the details of the electro-weak symmetry breaking sector and,
eventually, of its extension, as well as to take into account the effects from the Wilson loop of the weak, and
eventually others, gauge bosons.

\subsection{3D vacua in Standard Model Extensions}
\label{app:ext3Dvacua}
Until now we restricted our discussion to the bare Standard Model action, dressed up just with
General Relativity, cosmological constant and neutrino masses. If there are new light d.o.f., which
for any reasons escaped direct and indirect search, the structure of the vacua may change dramatically.
Let us rapidly discuss some of the possibilities. 
Clearly sterile neutrinos, light scalars interacting gravitationally or vector fields with very small couplings would
have important effects on the analysis. String theory, and in general extra-dimensional theories,
usually produce, after the stabilization of the moduli, a plethora of light scalar fields, 
which interacts mainly gravitationally. The presence of such fields
may alter the form of the radion potential, removing, for instance, the neutrino minima and/or
creating new minima at higher scales.

More interesting would be the presence of an axion ``$a$". Indeed, besides the usual Casimir contribution,
its shift symmetry could be used to switch on a flux for its field strength along the compact dimension 
$$\oint_{S_1} d a=f\,.$$
In this way the following extra contribution to the effective potential would arise
\beq
\frac{f^2\,r^3}{4\pi R^4} \,. \nn
\eeq
In general the shift symmetry of the axion is broken to a discrete subgroup, which quantizes
the flux $f=n f_a$ in units of the axion decay constant $f_a$.
For the QCD axion, if it exists, cosmological bounds set $f_a\sim 10^{9}\div 10^{12}$~GeV (see e.g \cite{pdg}).
In this case a non-trivial flux would wipe out all Casimir vacua with $R_0\gtrsim f_a^{-1}$,
while vacua with $R_0$ smaller than this scale will remain because Casimir energy dominates over the flux in this region. 
Moreover, each of the surviving vacuum will be replicated $n$ times, 
with $n\sim 1/(R_0 f_a)$, each of the replica with a different flux label.
Besides the fluxes, also the Casimir contribution from the QCD axions can lead to interesting consequences. 
Notice indeed that the actual limits on the QCD axion mass are $m_a\sim 10^{-6}\div 10^{-2}$~eV \cite{pdg},
right on the neutrino-vacuum scale. The presence of an axion in this range 
would increase the probability to find a vacuum also in the case of Dirac neutrino.
Indeed with normal hierarchy the bounds on the lightest Dirac neutrino in order to have a local minimum 
are weakened to $m_{\nu_1}\gtrsim (4.85\div 6.4)\cdot 10^{-3}$~eV,
while with inverted hierarchy an AdS minimum would always exist, also in the Dirac case.
 
Another possible source of modification of the effective radion potential could be supersymmetry (SUSY).
At low energy a light goldstino or gravitino would clearly affect the structure of the minima.
The fact that at high energies the fermionic and bosonic d.o.f. are the same because of SUSY
suggests the possibility to have new vacua at the SUSY-breaking scale. Moreover 
at higher energies all the contributions to the Wilson loop would disappear again leaving 
a number of approximate moduli in the effective theory.

Finally, even without going out of the SM there are other ingredients that can be used to find
other vacua, like modifying the boundary conditions with discrete and/or continuous global symmetries
like the fermionic ${\mathbb Z}_2$ symmetry, $B-L$\dots\ or by considering compactifications on more than one dimension.
The latter possibility will be explored in the next sections. 
%
%
%
%
%
%
%
%
%
%
As we will see the analysis will be a little subtler than in the case of toroidal compactifications in higher-dimensional models, 
for in our low-dimensional setups several degrees of freedom will not be dynamical. 
We want to see what is the analogue of looking for minima in the radion potential for finding (meta)stable vacua.

\subsection{2D SM vacua}\label{2Dvacua}
\begin{figure}[t!]
\begin{center}
\includegraphics[width=14cm]{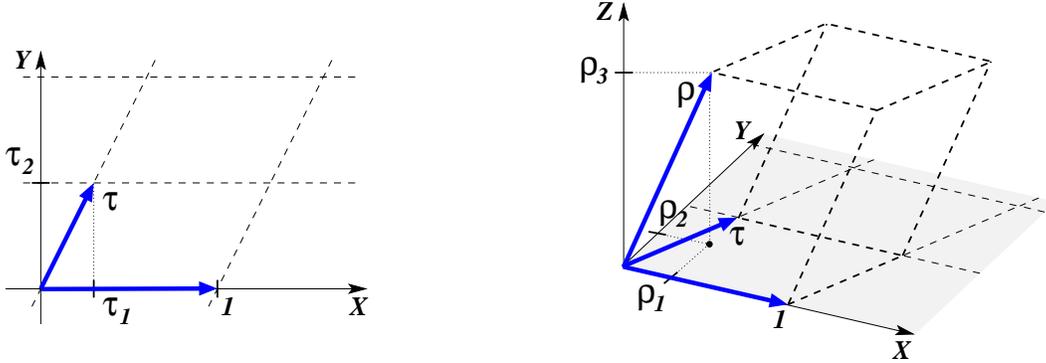}
\caption{\label{tori} \it \small {\em Left:} Standard parameterization of a two-torus. 
The spatial metric is flat in the $X$-$Y$ plane. The torus is defined by identifying points 
that differ by integer combinations of the two blue vectors, $(1,0)$ and $(\tau_1,\tau_2)$. 
{\em Right:} Adding a third dimension $Z$ and another vector $\rho$ gives an analogous parameterization for a three-torus.}
\end{center}
\end{figure}

Let us start by compactifying two spatial dimensions on a two-torus. The torus can be parameterized as usual by the area $A$ and by the complex modulus $\tau = \tau_1 +i \tau_2$, see fig.~\ref{tori}.
Calling $X^2$ and $X^3$ the compact coordinates with periodicity 1, $X^i \sim X^i + 1$, the metric takes the form \cite{PP}
\be
ds^2 = g_{\alpha\beta} d x ^\alpha d x ^\beta + A \,  \gamma_{ij} dX^i dX^j \; ,
\ee
with $i$ and $j$ labeling the compact directions, $\alpha$ and $\beta$ labeling the non-compact directions,  and
\be
\gamma_{ij} = \frac{1}{\tau_2} \left( 
\begin{array}{cc}
1 & \tau_1 \\
\tau_1 & |\tau|^2
\end{array}
\right) \; .
\ee
Notice that $\gamma_{ij}$ has unit determinant, so $A$ really measures the area of the torus. For simplicity we are not including the ``graviphotons'' $g_{\alpha 2}$ and $g_{\alpha 3}$ in the metric above---their inclusion would not qualitatively change the picture.

We dimensionally reduce the 4D Einstein-Hilbert action by imposing that the fields $g_{\alpha\beta}$, $A$, and $\tau$ only depend on the non-compact coordinates $(t,x)$. We get
\be \label{S2}
S =  \int \! d^2 x \sqrt{-g_{(2)}} \, \bigg[ \sfrac12 M_4^2 \big(
A \, {\cal R}_{(2)} + \frac{A}{2 \tau_2^2} |\di_\alpha \tau|^2\big)
- V (A, \tau)  \bigg] \; ,
\ee
where we included a potential energy for $A$ and $\tau$, coming from diverse sources like those studied throughout the paper.

Notice that if we imagine starting from 3D rather than 4D and compactifying one dimension on a circle, we end up with the same action as above 
for the radion field $R$,
\be
S  = \int \! d^2 x \sqrt{-g_{(2)}} \, \big[ \sfrac12 M_3 \,
R \, {\cal R}_{(2)} - V (R)  \big] \; ,
\ee
and obviously no $\tau$ degrees of freedom. 
Therefore everything we say in this section is readily exportable to this case as well, 
and in particular it applies to the 3D $ \to $ 2D  interpolations discussed in sect.~\ref{quantumhorizons}.

\subsubsection*{General analysis}

Like the radion in a compactification from 4D to 3D, $A$  does not have a kinetic term on its own. But unlike in the radion case, we now cannot demix
$A$ from the 2D metric by means of a suitable conformal transformation. 
To see this and in order to get some intuition on the dynamics of the system, it is instructive to work in $D=2+\epsilon$ dimensions.
Then the needed conformal transformation for going to Einstein frame is 
\be
g_{\alpha\beta} = A^{-2/\epsilon} \hat g_{\alpha\beta} \; .
\ee
This demixes $A$ from the metric and generates a kinetic term for $A$. The action becomes\
\be
S_{2+\epsilon} = \int \! d^2 x \sqrt{- \hat g_{(2+\epsilon)}} \,  \bigg[ 
\hat {\cal R}_{(2+\epsilon)} + \frac{1}{\epsilon} \, \frac{(\di_\alpha A)^2}{A^2} + \frac{|\di_\alpha \tau|^2}{2 \tau_2^2} 
- A^{-(2+\epsilon)/\epsilon} V (A, \tau) \bigg]  \; ,
\ee
where for notational convenience we set $\sfrac12 M_4^2=1$.
This procedure is obviously singular for $\epsilon \to 0$, but the divergence of the $A$
kinetic term suggests that in $D=2$ fluctuations of  $A$ are decoupled. Indeed 
the canonically normalized area field $\phi = \frac{1}{\sqrt \epsilon} \log A/A_0$ becomes a free field when we send $\epsilon$ to zero,
\be
{\cal L}_\phi = (\partial \phi)^2 - A_0 ^ {-\frac{2+\epsilon}{\epsilon}} e^{-\frac{2+\epsilon}{\sqrt{\epsilon}} \phi } \cdot V(e^{\sqrt \epsilon  \phi}, \tau)
\quad \to \quad (\partial \phi)^2 \; ,
\ee
and the corresponding fluctuations in the area vanish, $A = A_0 \, e^{\sqrt \epsilon \phi} \to A_0$.

The `vacua' are the minima of the effective potential $V_{\rm eff} = A^{-(2+\epsilon)/\epsilon} V$,
which correspond to points where $\di_{\tau_{1,2}} V= 0$ and
\be
\frac{2+\epsilon}{\epsilon} V = A \,  \di_A V \; .
\ee
In the $\epsilon \to 0$ limit vacua are characterized by a vanishing $V$.
Einstein's equations give us the curvature of these vacua,
\be
\hat G_{\alpha\beta} = - V_{\rm eff} \, g_{\alpha\beta}  \qquad
\Rightarrow \qquad \hat {\cal R}_{(2+\epsilon)} = \frac{2+\epsilon}{\epsilon} 
A^{-(2+\epsilon)/\epsilon} V = 
A^{-2/\epsilon}  \di_A V
\; .
\ee
Although the Einstein-frame curvature goes to zero for vanishing $\epsilon$, the curvature in the original conformal frame is finite, ${\cal R}_{(2)} = \di_A V$. So it is $\di_A V$ rather than $V$ itself that plays the role of a cosmological constant.

Notice that for small but finite $\epsilon$  the curvature of the effective potential on vacuum solutions is
\be
\partial_A ^2 V_{\rm eff} = A^{-(2+\epsilon)/\epsilon} \bigg[ -\frac{(6+2 \epsilon)}{\epsilon} \, \frac{\partial_A V}{A} + \partial^2 _A V \bigg]
\simeq - \frac{6}{\epsilon} A^{-2/\epsilon} \cdot \partial_A V \; .
\ee
Since de Sitter vacua correspond to positive $\partial_A V$ while Anti-de Sitter ones have negative $\partial_A V$, we conclude that the former sit at maxima of the effective potential and are therefore unstable, while the latter are stable minima. This is to be contrasted with the higher-dimensional cases, where dS/AdS vacua can be either local maxmima or minima. Of course when we send $\epsilon$ to zero the area $A$ decouples---as we saw the canonically normalized field sees no potential at all in the $D=2$ limit---and the de Sitter vacua are stable as well. Still this could be the formal reason why in interpolating from 3D to dS$_2$ one ends up with solutions that have the interpretation of ``top of the hill'' inflation, as we found in sect.~\ref{todSsec}.

The analysis we just sketched gives indeed the correct results, as we will now see by analyzing directly the action (\ref{S2}).
We first want to characterize the vacua. Vacuum solutions are solutions in which the moduli $A$ and $\tau$ have constant vev's, so we set to zero all gradients in the field equations.
$A$ appears in the action (\ref{S2}) as a Lagrange multiplier. Variation with
respect to $A$ yields the equation
\be \label{R_2}
{\cal R}_{(2)} = \di_A V \; ,
\ee
which determines ${\cal R}_{(2)}$.
The metric too is non-dynamical in 2D. In fact the variation of ${\cal R}_{(2)}$ is a total derivative, and Einstein's equations are thus a constraint on the matter sector,
\be
V = 0 \; .
\ee
Finally, variation with respect to $\tau_{1,2}$ yields a standard stationarity condition,
\be
 \di_{\tau_{1,2}} V= 0 \; .
\ee 
We thus see that 2D vacua are points in the $A$-$\tau$ space in which $V$ vanishes and is stationary with respect to $\tau$. Then $\di_A V$ determines the effective two-dimensional c.c., through eq.~(\ref{R_2}).

We now study the stability of such vacua against small fluctuations.
For simplicity we consider just the Minkowski case, that is we assume $\di_A V = 0$ on the vacuum solution. The analysis can be easily extended to the dS and AdS cases. We perturb the vacuum with small fluctuations $h_{\alpha\beta}$, $\delta A$, $\delta \tau_{1,2}$. The linearized field equations coming from varying the action with respect to $A$, $g_{\alpha\beta}$, and $\tau$ read respectively
\bea
\delta {\cal R}_{(2)} - m^2_{AA} \, \delta A - m^2_{A \tau_a} \,  \delta \tau_a & = & 0  \label{dR} \\
\di_\alpha \di_\beta \delta A -  \eta_{\alpha \beta} \, \Box \delta A & = & 0 \label{dA} \\
- c \, \Box \delta \tau_a - m^2_{\tau_a A} \, \delta A  - m^2_{\tau_a \tau_b}  \, \delta \tau_b & = & 0  \label{dtau}
\eea
where the mass matrix $m^2$ is given by $m^2_{\phi_i \phi_j} = \di_{\phi_i} \di_{\phi_j} V$, $c$ is the combination $A/\tau_2 ^2$  evaluated on the vacuum, and we made use of the vacuum equations above. Also all contractions are done by means of the background metric $\eta_{\alpha\beta}$.
Eq.~(\ref{dR}) is not a propagation equation for the metric, but rather it is a constraint.
This is because in two dimensions the Ricci and Riemann tensors are both determined by the Ricci scalar ${\cal R}_{(2)}$, with the appropriate tensor structures given by the metric.
Thus the only invariant quantity is ${\cal R}_{(2)}$, and knowing ${\cal R}_{(2)}$ uniquely determines the metric up to gauge transformations. Therefore eq.~(\ref{dR}) fixes the metric as a function of the other fields.
Eq.~(\ref{dA}) is not dynamical either. In fact its trace imposes $\Box \delta A = 0$, which plugged back into eq.~(\ref{dA}) itself gives
\be
\di_\alpha \di_\beta \delta A = 0 \; .
\ee
This constrains $\delta A$ to be a linear function of $x^\alpha$; in particular no localized perturbation can be given as an initial condition for $\delta A$. In studying the stability of the system against local perturbations we thus have to set $\delta A = 0$. We are then left with eq.~(\ref{dtau}) with $\delta A = 0$, which describes a stable system if and only if the mass matrix $m^2_{\tau_a \tau_b}$ is positive definite. In conclusion, in order for a 2D vacuum to be stable it must be a minimum of the potential along the $\tau_1$, $\tau_2$ directions.

\subsubsection*{The real world}

So far our discussion has been completely general. We now consider the case of the Standard Model with a small cosmological constant $\Lambda_4$.
Minimally, the two-dimensional potential has a positive contribution from the c.c.,
\be
V_\Lambda = \Lambda_4 \, A \; ,
\ee
as well as negative and positive contributions from the Casimir energy of bosons and fermions, respectively.
The computation of the Casimir energy on the torus proceeds analogously to the cylinder case of Appendix \ref{app:Casimir}:
one writes the two-point function as a sum over images and subtracts the UV divergent part; then the Casimir $T_{\mu\nu}$ is just given by proper derivatives of the resulting Green's function in the limit where the two points are brought together. The result in 4D for a massless boson is of the form (see e.g.~ref.~\cite{PP})
\be
\rho_{\rm Casimir} \propto  - \frac{1}{A^2} {\sum_{n,m} }' \frac{1}{| n-m \tau |^4} \; ,
\ee
where the primed sum extends from $-\infty$ to $+\infty$ excluding the case $(n,m)=(0,0)$. After dimensional reduction this gives a contribution to the two-dimensional potential $V_{\rm Casimir} = \rho _{\rm Casimir} A$.
For massless fermions the result is the same, apart from the overall sign which is positive. For massive particles the result is obviously much more complicated, and we cannot simply model it with a step function like in the cylinder case: for instance for a square torus we could say that a massive particle does not contribute to the Casimir energy until the area $A$ drops below $1/m^2$, and after that it contributes like a massless particle, but for a general torus the combined dependence of this threshold on $A$ and $\tau$ will be more involved.

We will not attempt here a detailed analysis of the Casimir energy in the SM as a function of the torus moduli in order to find stable 2D vacua. 
There is however a simple situation that we can readily study. Suppose that starting from 4D we first compactify $z$ stabilizing the radius at $R_z \sim 1 \: {\rm mm}$, thanks to the interplay between the c.c.~and the Casimir energy as described in sect.~\ref{sec:SMlandscape}. Then we compactify another dimension, say $y$, on a much larger circle, so that we can consistently use the 3D effective theory. We want to see if in this situation we can find a stable vacuum. In the 3D theory we have a cosmological constant $\Lambda_3$ and, among other things, a massless photon. We can then turn on a constant electric field along the (non-compact) $x$ direction, $E = F_{0x}$; alternately we can turn on an electric field for the graviphoton. Such an electric field does not break 2D Lorentz invariance in the non-compact dimensions, since $F_{\alpha\beta} \propto \epsilon_{\alpha\beta}$ is Lorentz-invariant.
Equivalently, in 3D a 1-form is dual to a scalar $\phi$, and a constant electric field along $x$ corresponds to a constant $\di_y \phi = E$. So the case we are studying is technically the lower dimensional analogue of the axion wrapped around the circle of sect.~\ref{app:ext3Dvacua}. The electric field gives a positive contribution to the 2D energy density that scales like $f^2 / R_z R_y$, where $R_y$ is the physical radius of the $y$ dimension and $f = E \cdot R_z R_y$ is the conserved flux. Then for large $R_y$ and $f$ the Casimir energy coming from the compactness of $y$ is completely negligible with respect to this  classical contribution to the potential.

The full 2D potential we consider is therefore
\be
V = \Lambda_3  R_y + \frac{ f^2  }{R_z R_y} \; .
\ee
Recall that 2D vacua are characterized by an overall vanishing potential energy. Therefore in the presence of a flux $f$ we only get a vacuum if the 3D cosmological constant is negative, which could well be the case for the SM as we argued in sect.~\ref{sec:SMlandscape}. In this case the radius is given by
\be
R_y = \frac{f}{\sqrt{R_z |\Lambda_3|}} = R_z \, 
\frac{f}{\varepsilon} \; ,
\ee
where we defined the quantity $\varepsilon = \sqrt{R_z^3 |\Lambda_3|}$.
Without fine-tunings of the neutrino masses we expect $\varepsilon$ to be
of order one. However as we will soon see for our approximations to be self-consistent we will have to assume $\varepsilon \ll 1$.
The curvature of this vacuum is determined by $\frac{2}{M_3} \di_{R_y} V$, where the 3D Planck mass is $M_3 = R_z \cdot M_4^2$; we thus get
\be
{\cal R}_{(2)} = \frac{2}{M_4^2} \, \frac{1}{R_z} \di_{R_y} V = -\frac{4}{M_4^2}   \frac{\varepsilon^2}{R_z^4} \,
\ee
independent of $f$. The flux $f$ is quantized in units of the 4D electric charge $e$ , so we have a {\em discretum} of different AdS$_2$ vacua parameterized by $f$, all with exactly the same curvature radius, which is roughly $1/\varepsilon$ times larger than our Hubble scale.

For our approximations to be self-consistent we first have to assume that $R_y \gg R_z$, which requires $f \gg \varepsilon$. Then we have to impose that the electric field does not destabilize the 3D radion, $R_z$: after all $R_z$ was stabilized thanks to the Casimir energy, which is a small quantum effect. $R_z$ is still stable if the ``force'' $\di_{R_z} V$ is smaller than the typical curvature scale of the stabilizing potential for $R_z$, so that the electric flux only moves $R_z$ slightly away from the minimum. Notice that $\Lambda_3$ is implicitly a function of $R_z$, but by assumption $R_z$ is at a minimum of $\Lambda_3$, so we get no force from that piece of the potential. We have
\be
\di_{R_z} V = - \frac{f^2}{R_z^2 R_y} =  - \frac{f \, \varepsilon}{R_z^3} \;, 
\ee
to be compared with $\sim 1/R_z^3$. We thus have to impose $f \ll 1/ \varepsilon$, which combined with the previous requirement, $f \gg \varepsilon$, tells us that our approximations are self consistent only if $\varepsilon$ is much smaller than one, i.e.~if the 3D cosmological constant is unnaturally small. Therefore in general we don't expect these very asymmetric compactifications to give rise to stable vacua---one should  instead consider more symmetrically shaped tori, for which a full analysis of the Casimir energy as a function of the torus moduli is necessary.
However if the required fine-tuning is fortuitously realized in the real world and $\varepsilon$ is actually very small, then there exist $N \sim 1/ (\varepsilon e)$ two-dimensional AdS vacua, parameterized by $f$, all with the same 2D curvature length.

\subsection{No 1D ``vacua"}\label{1Dvacua}
We now imagine to compactify all three spatial dimensions on a three-torus. For the discussion that follows the parameterization we choose for the torus is not important, but for concreteness let us parameterize it with the overall (dimensionful) scale factor $a$ and the shape moduli $\Phi = (\tau_1, \tau_2, \rho_1, \rho_2, \rho_3)$ defined in fig.~\ref{tori}. Then if the compact directions have periodicity $1$, the 4D metric reads
\be
ds^2 = -N^2 dt^2 + a^2 \gamma_{ij} dX^i dX^j \; ,
\ee
with
\be
\gamma_{ij} = \frac{1}{(\rho_3  \, \tau_2 )^{2/3}} \left( 
\begin{array}{ccc}
1 & \tau_1 & \rho_1\\
\tau_1 & \tau_1^2 + \tau_2^2 & \rho_1 \tau_1+\rho_2 \tau_2 \\
\rho_1 & \rho_1 \tau_1+\rho_2 \tau_2 &  \rho_1^2 + \rho_2^2+\rho_3 ^2
\end{array}
\right) \; .
\ee
Notice that $\det \gamma_{ij} =1$ as before, so the volume of the torus is $a^3$. Dimensional
reduction yields the 1D action
\be \label{1Daction}
S = \int \! dt \, \sfrac12 M_4^2 \bigg[ -\frac{6 \,\dot a^2 a}{N} + \frac{a^3}{N} \,
\dot \Phi \cdot K(\Phi) \dot \Phi \bigg] - N \, V(a, \Phi) \; ,
\ee
where $V$ is the sum of the 4D cosmological constant, Casimir energy density, and possibly other sources of potential energy, all multiplied by the
volume of the three-torus, and $K(\Phi)$ is a 5$\times$5 matrix that depends on the shape of the torus and whose explicit form we spare the reader. Its positivity can be readily checked for very symmetric configurations, like the rectangular torus $\tau_1= \rho_1 = \rho_2 = 0$. 
More generically, the $\Phi$'s parameterize the coset manifold $SL(3)/SO(3)$, and being $SO(3)$ the maximal compact  subgroup of $SL(3)$, the corresponding non-linear sigma model has positive definite kinetic energy.

Generically the action (\ref{1Daction}) describes a cosmology. $N$ appears as a Lagrange multiplier and its equation of motion is the Hamiltonian constraint ${\cal H}=0$, which is nothing but the Friedman equation 
\be \label{friedmann1}
{\cal H} =  \sfrac12 M_4^2 \big[ -6 \,\dot a^2 a + a^3 \,
\dot \Phi \cdot K(\Phi) \dot \Phi \big] + V(a,\Phi)= 0 \; ,
\ee
where we fixed the gauge $N=1$. We can equivalently set $N=1$ directly in the Lagrangian,
\be \label{lag}
{\cal L} =   \sfrac12 M_4^2 \big[ - 6 \, \dot a^2 a+ a^3 \,
\dot \Phi \cdot K(\Phi) \dot \Phi \big] -  V(a, \Phi) \; ,
\ee
and supplement the system by the constraint that the total Hamiltonian vanishes, eq.~(\ref{friedmann1}). Since the Hamiltonian is conserved on the equations of motion of ${\cal L}$, this is just a constraint on the initial conditions. Notice that $a$ enters the action with negative kinetic energy.

Usually cosmological solutions evolve with time. In our case this time evolution would correspond to a decompactification, or to a big crunch, or to some anisotropic Kasner-like solution. Instead we are looking for vacua---i.e.~static solutions in which the moduli are stabilized. It is evident from the Lagrangian above that a static solution must be an extremum of the potential, but then
to have zero total energy the potential itself should vanish  at the same point. So the existence of a truly static solution requires a perfectly tuned potential.
 
More realistically, in our case we expect $V$ to develop non trivial features at the micron scale with typical energies of order $\mu$m$^{-1}$. So let us assume that there is a stationary point of $V$ at $a_0 \sim {\rm \mu m}$, $\Phi_0 \sim 1$, 
but for reasons that will soon become clear let's assume that the potential itself at the stationary point is somewhat smaller than the typical energy scale:
$V_0 \sim \varepsilon^2 \; {\rm \mu m}^{-1}$ with $\varepsilon \ll 1$. Then in a neighborhood of $(a_0, \Phi_0)$ we can expand the Lagrangian at second order in the displacements $\delta a, \delta \Phi$. The resulting quadratic Lagrangian describes a set of harmonic oscillators, provided that the Hessian of the potential---the `mass matrix'---have the right signature. In particular, since the $\Phi$'s have a positive definite kinetic energy while that of $a$ is negative,
$V$ should be positively curved along the $\Phi$ directions and negatively curved along $a$ \footnote{This is true if the mixed second derivatives $\di_a \di_\Phi V$ are negligible. The general condition for having only oscillatory solutions is that the Hessian of $V$ with respect to $\Phi$ and to $i \cdot a$ be positive definite.}. If these conditions are met, then the typical oscillation frequency in all directions is of order of our present Hubble rate $H_0$---assuming that
the curvature scales of $V$ are of order $\mu$m$^{-1}$. 

The Hamiltonian constraint fixes the initial oscillation amplitudes. The total energy of the oscillators should vanish, taking into account also the offset $V_0\sim \varepsilon^2 \; {\rm \mu m}^{-1}$.  So the typical amplitudes are $\delta a \sim \varepsilon \; {\rm \mu m}$, $\delta \Phi \sim \varepsilon$. If $\varepsilon$ is small the oscillations are small compared to the typical variation scales of the potential, and the perturbative analysis we are sketching here is justified. In this case we have an almost static micron-sized universe that undergoes small periodic oscillations in size and shape on a timescale of order $10^{10}$ years! Note that a classical description 
of this motion is justified since the amplitudes of oscillation are much larger than the quantum uncertainties, with $(\delta_{quantum} a/\delta_{classical} a)^2 \sim \mu{\rm m} \times H_0$. 

Of course the fact that $\delta a$ has negative energies signals that the system is unstable once interactions between the two sectors---the ``inverted'' oscillator and the normal ones---are taken into account. The two sectors can start exciting each other while keeping the total energy fixed, and this happens classically already at  perturbative level. However unlike in relativistic field theories with ghosts where the rate of such instability is formally infinite because of Lorentz symmetry, here the instability is slow and its rate can be reliably computed in perturbation theory.

Despite the appearance of the Planck scale in front of the kinetic terms, the only suppression of interactions in our case comes from the smallness of $\varepsilon$---i.e.~interactions are not Planck-suppressed. This is because we are studying large classical oscillations, much larger than the typical quantum spread of the ground-state wave-function. Then in the Lagrangian~(\ref{lag}) we can reabsorb $M_4$ into a redefinition of time. This only changes the overall normalization of the action, which classically is arbitrary. With this redefinition, there is no small parameter in the Lagrangian, and the importance of interactions is only controlled by the oscillation amplitude, $\varepsilon$.
Therefore the instability rate is suppressed with respect to the oscillation frequency by positive powers of $\varepsilon$,
\be
\Gamma \sim H_0 \, [ \varepsilon+ \varepsilon^2 + \dots] \; .
\ee
It is easy to convince one's self that the leading term is there only if resonance phenomena are possible, i.e.~if two frequencies are tuned to be equal.
Barring this possibility, the instability rate is generically of order $\Gamma \sim \varepsilon^2 H_0$. 
A detailed analysis of the classical dynamics of two coupled harmonic oscillators, one of which has negative energy, confirms this quick estimate.

In conclusion, if the potential energy at the stationary point is much smaller than the typical energy scales and the mass matrix has the right signature, than there exists a micron sized solution that slightly oscillates in size and shape with a period of order $H_0^{-1}$. Eventually it is unstable against decompactification or crunching, but on a longer timescale of order $H_0^{-1} / \varepsilon^2$.
If instead there are no special tunings in the potential, then the instability time is of order of the would-be oscillation frequency and there is no conceptual difference between the situation we are describing and a standard cosmological solution.


\end{document}